\PassOptionsToPackage{caption=false,font=footnotesize}{subfig}
\documentclass[conference]{IEEEtran}

\usepackage{amsmath,amssymb,amsfonts}
\usepackage{graphicx}
\usepackage{textcomp}
\usepackage{booktabs}
\usepackage{multirow}
\usepackage{subcaption}
\usepackage{microtype}
\usepackage{url}
\usepackage[utf8]{inputenc}
\usepackage[T1]{fontenc}
\usepackage{algorithm}
\usepackage{algpseudocode}
\usepackage{siunitx}
\usepackage[table,xcdraw,dvipsnames]{xcolor}
\usepackage{float}
\usepackage{subfig}  

\usepackage[style=ieee]{biblatex}
\usepackage{pifont}
\newcommand{\cmark}{\ding{51}}
\newcommand{\xmark}{\ding{55}}

\begin{document}

\title{RegimeFolio: A Regime Aware ML System for Sectoral Portfolio Optimization in Dynamic Markets}

\author{
\IEEEauthorblockN{Yiyao Zhang\IEEEauthorrefmark{1},
Diksha Goel\IEEEauthorrefmark{2},
Hussain Ahmad\IEEEauthorrefmark{1},
and Claudia Szabo\IEEEauthorrefmark{1}}
\IEEEauthorblockA{\IEEEauthorrefmark{1}The University of Adelaide, Australia\\
\texttt{yiyao.zhang@student.adelaide.edu.au};
\texttt{\{hussain.ahmad, claudia.szabo\}@adelaide.edu.au}}
\IEEEauthorblockA{\IEEEauthorrefmark{2}CSIRO's Data61, Australia\\
\texttt{diksha.goel@csiro.au}}
}

\maketitle

\begin{abstract}
Financial markets are inherently non-stationary, with shifting volatility regimes that alter asset co-movements and return distributions. Standard portfolio optimization methods, typically built on stationarity or regime-agnostic assumptions, struggle to adapt to such changes. To address these challenges, we propose RegimeFolio, a novel regime-aware and sector-specialized framework that, unlike existing regime-agnostic models (e.g., DeepVol, DRL optimizers), integrates explicit volatility regime segmentation with sector-specific ensemble forecasting and adaptive mean–variance allocation. This modular architecture ensures forecasts and portfolio decisions remain aligned with current market conditions, enhancing robustness and interpretability in dynamic markets. RegimeFolio combines three components: (i) an interpretable VIX-based classifier for market regime detection; (ii) regime and sector-specific ensemble learners (Random Forest, Gradient Boosting) to capture conditional return structures; and (iii) a dynamic mean–variance optimizer with shrinkage-regularized covariance estimates for regime-aware allocation. We evaluate RegimeFolio on 34 large cap U.S. equities from 2020 to 2024. The framework achieves a cumulative return of 137\%, a Sharpe ratio of 1.17, a 12\% lower maximum drawdown, and a 15–20\% improvement in forecast accuracy compared to conventional and advanced machine learning benchmarks. These results show that explicitly modeling volatility regimes in predictive learning and portfolio allocation enhances robustness and leads to more dependable decision-making in real markets.
\end{abstract}


\section{Introduction}
Financial markets are increasingly unstable, with abrupt regime shifts triggered by macroeconomic shocks, geopolitical tensions, policy changes, and shifts in investor behaviour \cite{da2015fear}. Events such as the COVID-19 pandemic, persistent inflationary pressures, and fluctuations in interest rate policy have highlighted the fragility of portfolio management techniques that assume stable statistical relationships \cite{ang2002asset}. In such conditions, cross-asset and cross-sector correlations often rise sharply, eroding diversification benefits at precisely the moments when they are most needed \cite{billio2012econometric}. While prior works, including DeepVol~\cite{ramponi2024deepvol} and DRL-based optimizers~\cite{masuda2024portfolio}, have advanced volatility forecasting and portfolio optimization, they remain regime-agnostic and often overlook sectoral heterogeneity. This leads to misaligned forecasts and unstable allocations when market conditions shift~\cite{guidolin2008asset, pastor2012are}. \textit{This emphasizes the need for adaptive investment strategies capable of operating effectively in non-stationary, volatility-driven environments.}

The advent of artificial intelligence has accelerated innovation across domains, including software engineering \cite{ahmad2025future, haque2022think, abdulsatar2025towards}, cybersecurity \cite{ahmad2025survey, chopra2024chatnvd, goel2024machine, ullah2025skills}, and cloud computing \cite{ahmad2025towards, ahmad2024smart, ahmad2025resilient}. Within finance, Machine Learning (ML) has advanced time-series modelling by capturing complex, non-linear patterns in asset return dynamics, offering greater flexibility and predictive power than traditional approaches. However, most existing ML-based approaches are regime agnostic, training on pooled historical data that mixes bull, bear, and transitional markets~\cite{gu2020empirical, d2022hidden}. This mixing of heterogeneous market states pollutes learned parameters and produces allocations that are disoriented with existing conditions. Moreover, many methods treat the market as a single homogeneous system, neglecting that different economic sectors respond differently to macroeconomic and volatility shocks~\cite{bekaert2009international, belo2014macroeconomic}. \textit{While recent deep learning and reinforcement learning approaches offer adaptability, they often lack interpretability, omit explicit modelling of temporal regime structure, and are vulnerable to performance deterioration under distributional shifts.}

To address these limitations, we propose \textbf{RegimeFolio}, a hierarchical, regime-aware machine learning framework for dynamic portfolio optimization. This framework unifies the following components: (i) explicit volatility regime segmentation using the   Chicago Board Options Exchange (CBOE) VIX, (ii) sector-specific ensemble forecasting models trained separately for each regime, and (iii) a regime conditioned mean–variance optimization (MVO) module that adapts allocations to the prevailing risk–return environment. The framework conditions both the forecasting and allocation stages on market regime and sector context, to align decision-making with the inherently non-stationary and heterogeneous structure of financial markets.

Beyond its architectural novelty, \textit{RegimeFolio} delivers distinct advantages over existing ML-based portfolio optimization approaches. First, by conditioning both forecasts and allocations on explicitly defined volatility regimes, it mitigates parameter contamination from pooled training data and enhances stability under distributional shifts. Second, the use of sector-specialized forecasters enables the framework to capture heterogeneous responses of industries to macroeconomic and volatility shocks, a dimension often overlooked by homogeneous or regime-agnostic models. Third, the modular forecast–then–optimize design preserves interpretability and computational efficiency, in contrast to end-to-end DRL pipelines that frequently sacrifice transparency for adaptability. Collectively, these benefits underpin the framework’s robustness, evidenced by reduced forecasting error, superior risk-adjusted performance across regimes, and smaller drawdowns during periods of heightened market stress.

Moreover, the importance of this approach resides in its architectural shift from static, regime blind prediction to a fully adaptive, volatility-sensitive allocation process. Unlike methods such as DeepVol~\cite{ramponi2024deepvol} or DRL-based optimizers~\cite{masuda2024portfolio}, which average across incompatible regimes or model the market as an undifferentiated vector space, \textit{our proposed RegimeFolio} explicitly decomposes the problem into macro regime recognition, sector-level signal learning, and context-aware allocation. This separation addresses the instability of asset behaviours across market phases and supports proactive, state-aware portfolio management, particularly during structural transitions when traditional models are most prone to failure.

\textit{We evaluate our proposed framework on daily data from 34 large cap U.S. equities over 2020–2024, covering multiple volatility regimes and market events. Empirical results show consistent gains in both predictive accuracy as well as risk adjusted returns, relative to regime-agnostic ML baselines. These findings showcase the potential of regime-aware, sector-specialized architectures to deliver scalable, interpretable, and practically deployable solutions for portfolio optimization in volatile and evolving markets.}

 \noindent \textit{\textbf{Our key contributions are:}}
 \begin{itemize}
     \item \textbf{Regime-aware portfolio optimization framework:} We propose RegimeFolio, a modular pipeline where volatility regime detection, sector-specific forecasting, and regime conditioned allocation are unified in one single pipeline, addressing the temporal non-stationarity and sectoral heterogeneity directly.

     \item \textbf{VIX-based regime segmentation:} We empirically validate a segmentation method using the CBOE VIX to classify market states, ensuring consistency across forecasting and allocation.

     \item \textbf{Empirical performance:} Exhaustive experiments on 34 large cap U.S. equities (2020–2024) show up to 20\% lower forecast error (MAE) and Sharpe ratio improvements exceeding 0.5 compared to regime-agnostic ML baselines and an S\&P 500 benchmark.

     \item \textbf{Quantification of regime sensitive sector dynamics:} Our experimental evaluation shows how sector betas and cross asset correlations vary across regimes, demonstrating that ignoring these shifts eventually leads to sub-optimal allocations and degraded forecasts.

     \item \textbf{Scalability, interpretability, and deployment readiness:} We demonstrate that the architecture trains quickly (in few minutes) on commodity hardware, supports parallel scaling across assets, and can be integrated into standard portfolio management workflows without sacrificing interpretability.\\
     
 \end{itemize}

 \noindent The remainder of this paper is organized as follows: Section 2 reviews the related literature; Section 3 outlines the foundational principles and design rationale; Section 4 details the phased research methodology; Section 5 presents the experimental setup, and results; Section 6 offers an in-depth discussion; and Section 7 concludes with key insights.

 \begin{table*}[!t]
 \centering

 \caption{Comparison of Forecasting and Structural Modeling Approaches}
 \label{tab:forecasting_models}
\renewcommand{\arraystretch}{1.2}
 
\scriptsize
 \begin{tabular}{llcccc}
 \hline
 \textbf{Study} & \textbf{Method} & \textbf{Regime Awareness} & \textbf{Sector Modeling} & \textbf{Volatility Modeling} & \textbf{Dynamic Allocation} \\
 \hline
 Ramponi et al.~\cite{ramponi2024deepvol} & Dilated Causal CNN & \xmark & \xmark & \cmark & \xmark \\

 Taylor et al.~\cite{taylor2023forecasting} & Deep Learning Ensemble & \xmark & \xmark & \cmark & \xmark \\

 Cho et al.~\cite{cho2025forecasting} & Kolmogorov–Arnold Network (KAN) & \xmark & \xmark & \cmark~(VIX) & \xmark \\

 Sonani et al.~\cite{li2021hierarchical} & Graph Neural Networks (GNN) & \xmark & \cmark & \xmark & \xmark \\

 \textbf{RegimeFolio (Our Work)} & {Regime Aware Hierarchical Optimization} & \textbf{\cmark~(Explicit)} & \textbf{\cmark} & \textbf{\cmark} & \textbf{\cmark} \\
 \hline
 \end{tabular}

 \end{table*}

 \begin{table*}[!t]
 \centering
 \caption{Comparison of Machine Learning Based Portfolio Optimization Approaches}
 \label{tab:portfolio_models}
\renewcommand{\arraystretch}{1.5}
 
\scriptsize
 \begin{tabular}{llcccc}
 \hline
 \textbf{Study} & \textbf{Method} & \textbf{Regime Awareness} & \textbf{Sector Modeling} & \textbf{Volatility Modeling} & \textbf{Dynamic Allocation} \\  \hline

 Masuda~\cite{masuda2024portfolio} & Ensemble Learning + Convex Optimization & \xmark & \xmark & \cmark & \cmark \\

 Chen et al.~\cite{wang2024drl} & Deep Reinforcement Learning (DRL) & $\sim$~(Implicit) & \xmark & $\sim$~(Implicit) & \cmark \\

 Wang and Zhou~\cite{wang2024deep} & Graph Enhanced DRL & $\sim$~(Implicit) & \cmark & $\sim$~(Implicit) & \cmark \\

 \textbf{RegimeFolio (Our Work)} & {Regime Aware Hierarchical Optimization} & \textbf{\cmark~(Explicit)} & \textbf{\cmark} & \textbf{\cmark} & \textbf{\cmark} \\
 \hline
 \end{tabular}

 \end{table*}

 \section{Related Work}

Advances in machine learning (ML), combined with the increasing availability of high-frequency financial data, have driven significant progress in portfolio modeling. This paper reviews three key domains: volatility regime modeling, market structure analysis, and ML-driven portfolio allocation, that collectively inform the development of our hierarchical, regime-aware framework. While each domain has contributed valuable insights, most approaches operate in isolation, limiting their robustness in dynamic, non-stationary market environments. By identifying and synthesizing these gaps, we establish the foundation for our integrated methodology.

 \subsection{Volatility Regime Modeling and Forecasting}

In financial time series, deep learning models have shown promising empirical success in volatility forecasting by seizing non-linear dependencies and high-frequency patterns. Ramponi et al.~\cite{ramponi2024deepvol} introduced \textit{DeepVol}, a dilated causal convolutional neural network that overtakes traditional GARCH-based models in performance. Likewise, Taylor et al. \cite{taylor2023forecasting} showcased that deep ensemble architectures are capable of generating accurate intra-day volatility distributions.

However, these models are typically trained over full historical datasets, assuming a stationary data generating process, implicitly. Which in turn overlooks the structural breaks and volatility regimes, such as transitions between crisis episode score bull and bear markets. Without explicit mechanisms for regime segmentation, such models are susceptible to \emph{domain shift} and often underperform during out of distribution (OOD) periods. Moreover, they face an inherent \textit{bias–variance trade off}: highly flexible models tend to overfit regime inconsistent signals, while constrained models may fail to capture dynamics unique to specific market states. As a result, their generalization capacity deteriorates under novel or extreme volatility conditions.

 In contrast, regime-aware classifiers, such as those used in financial distress prediction, have shown promise. For example, Liu et al.~\cite{liu2024ensemble} trained ensemble models to distinguish between “healthy” and “distressed” firms. However, these models are optimized for discrete classification and do not extend naturally to continuous volatility estimation. This bifurcation, between regime-agnostic volatility forecasters and regime-sensitive but narrowly scoped classifiers, remains a key limitation in financial time series modeling.

 \subsection{Modeling Market Structure}

 Financial markets are inherently interconnected, with sectoral and asset-level dependencies driving contagion and co-movement. A prominent approach for modeling these relationships involves Graph Neural Networks (GNNs), where nodes represent assets and edges encode economic or statistical links. Sonani et al.~\cite{li2021hierarchical} proposed a hybrid LSTM-GNN model that improves stock price prediction by integrating temporal sequence analysis with graph-based structural information. Similarly, Li et al.~\cite{li2024gnn} demonstrated that incorporating asset-level relational information enhances predictive performance during turbulent periods. While GNNs capture complex structural dependencies, they typically lack temporal segmentation. Embeddings are learned over the entire training set, encoding average structural patterns that may become unstable across different volatility regimes.

 The CBOE VIX, a well-established proxy for market sentiment and systemic risk, offers a complementary signal for temporal segmentation. Studies such as Schrimpf et al.\cite{schrimpf2018anatomy} show that VIX spikes align with macroeconomic disruptions. Recent architectures like Kolmogorov–Arnold Networks (KANs) have further improved VIX forecasting accuracy~\cite{cho2025forecasting}. Our framework provides an alternative to implicit GNN-based learning by using an explicit, two pronged approach to market structure: (1) sector-level disaggregation to model industry-specific behaviors and (2) regime conditioned covariance matrices to capture time-varying, market-wide dependencies. This approach makes the structural assumptions transparent and directly tied to an observable market indicator (VIX).

 \subsection{Machine Learning for Portfolio Optimization}

 ML-based portfolio optimization approaches can be broadly categorized into following two categories:

 \subsubsection{Forecast Then Optimize Architectures.} These frameworks, to which our work belongs, typically use ML models to predict expected returns and risk measures, which are then passed to classical optimization solvers~\cite{masuda2024portfolio}. Although this modular design supports interpretability and flexibility, the disjoint between prediction and allocation can result in \textit{prescription misalignment} if the models are not carefully aligned. Our framework mitigates this by conditioning both the forecast and risk models on the same regime signal.
  \vspace{0.1in}

 \subsubsection{End-to-End Deep Reinforcement Learning.} To address misalignment, several studies model portfolio optimization as a sequential decision-making problem using deep reinforcement learning (DRL). DRL agents learn to generate asset weights directly by optimizing a reward function like the Sharpe ratio~\cite{wang2024drl}. While adaptive, early DRL approaches often processed financial inputs as flat feature vectors, ignoring market structure.
  \vspace{0.1in}

 \subsubsection{Graph Enhanced DRL.} A significant advancement has been the integration of GNNs into DRL policy networks. Wang and Zhou~\cite{wang2024deep} proposed a GNN-enhanced DRL agent where asset relationships are encoded dynamically. However, these models often remain regime agnostic during training and can be opaque, reintroducing vulnerability to structural breaks.\\

Across volatility forecasters, structural graph models, and DRL optimizers, persistent limitations remain: regime-agnostic training that induces parameter contamination, structural representations that remain temporally static, and decision policies that lack transparency. These shortcomings underscore the need for a deeper investigation into the research gap in ML-driven portfolio optimization.

\noindent \textbf{Research Gap.} Tables~\ref{tab:forecasting_models} and~\ref{tab:portfolio_models} summarize the capabilities of recent forecasting and allocation approaches. Despite advances in model architectures and data availability, a persistent fragmentation remains: most methods are designed to address a single dimension of the problem in isolation. This narrow scope undermines \textit{robustness}, which we define as the ability of portfolio systems to deliver stable and reliable performance under distributional shifts—such as volatility spikes, structural breaks, and regime transitions.

In practice, pooled training contaminates model parameters across incompatible regimes, static structural approaches fail to capture the time-varying reconfiguration of cross-asset dependencies, and allocation models without explicit regime awareness exhibit instability when market conditions change. Collectively, these limitations lead to fragile portfolio systems that degrade precisely when adaptability and resilience are most critical. Specifically:

\begin{itemize}
    \item \textit{Volatility forecasters} capture temporal patterns in returns but neglect structural dependencies between assets and do not account for regime transitions.
    \item \textit{Structural models} (e.g., graph-based approaches) represent cross-asset relationships but are trained on pooled data, yielding temporally static embeddings that ignore regime-specific dynamics.
    \item \textit{Deep reinforcement learning (DRL) agents} adapt allocations dynamically but typically operate without explicit regime signals, and their opaque policies complicate interpretation and validation.
\end{itemize}

This persistent separation between temporal, structural, and decision-making components reduces the resilience of ML-driven portfolio optimization, especially during volatility spikes, structural breaks, and regime transitions. These are precisely the conditions under which investors require methods that are not only adaptive but also context-aware, able to recognize volatility regimes, capture sector-specific dynamics, and translate this knowledge into stable, interpretable allocation decisions. The absence of such integration leaves current approaches vulnerable to performance breakdowns at the moments of greatest market stress, underscoring the need for robust, regime-aware, and sector-sensitive frameworks.

\textit{We address this gap with a hierarchical, regime-aware \emph{predict then optimize} framework named \textbf{RegimeFolio} that implants volatility context consistently across all stages of the modeling pipeline. The approach begins with VIX-based regime segmentation, enabling the use of regime-specific forecasting models. It then applies sector-specialized ensemble forecasters to capture the distinct return dynamics of each sector within each regime. Finally, a regime-conditioned dynamic allocation module translates these forecasts into portfolio weights using forward-looking returns and a shrinkage-regularized, regime-specific covariance matrix. By integrating regime identification, sector-level modeling, and context-aware allocation in a unified architecture, the framework moves beyond monolithic predictors toward adaptive, interpretable decision systems suited to dynamic market environments. In contrast to existing ML-based approaches, RegimeFolio mitigates parameter contamination from pooled training, captures heterogeneous sectoral responses often overlooked by homogeneous models, and preserves interpretability and computational efficiency relative to black-box DRL pipelines. These advantages underpin its robustness, enabling consistent performance across volatility regimes, reduced forecast error, and smaller drawdowns under market stress.}

 \section{Foundational Principles and Design Rationale}

 Our \textit{RegimeFolio} framework is grounded in three complementary theoretical pillars: Modern Portfolio Theory (MPT), regime switching models, and sectoral heterogeneity in asset behaviour. Together, these provide the foundation for a volatility-sensitive, sector-specialized allocation mechanism capable of adapting to dynamic market conditions.

\subsection{Modern Portfolio Theory and  Role of Input Quality}
MPT formalises portfolio construction as the maximisation of expected return for a given level of risk through mean-variance optimization ~\cite{markowitz1952portfolio}. While MPT offers a mathematically elegant framework, its practical performance is highly sensitive to the accuracy of input estimates for returns, volatilities, and correlations \cite{maheu2000identifying, guidolin2008size}. In non-stationary financial markets, such estimates degrade rapidly, leading to unstable and sub-optimal allocations. 

Our design addresses this limitation by replacing static, full-sample estimates with regime-specific, dynamically updated inputs derived from ensemble machine learning forecasts. By localising parameter estimation to homogenous market regimes, we reduce estimation error and improve optimization stability.

\begin{figure*}
 \centering
 \includegraphics[width=0.8\paperwidth]{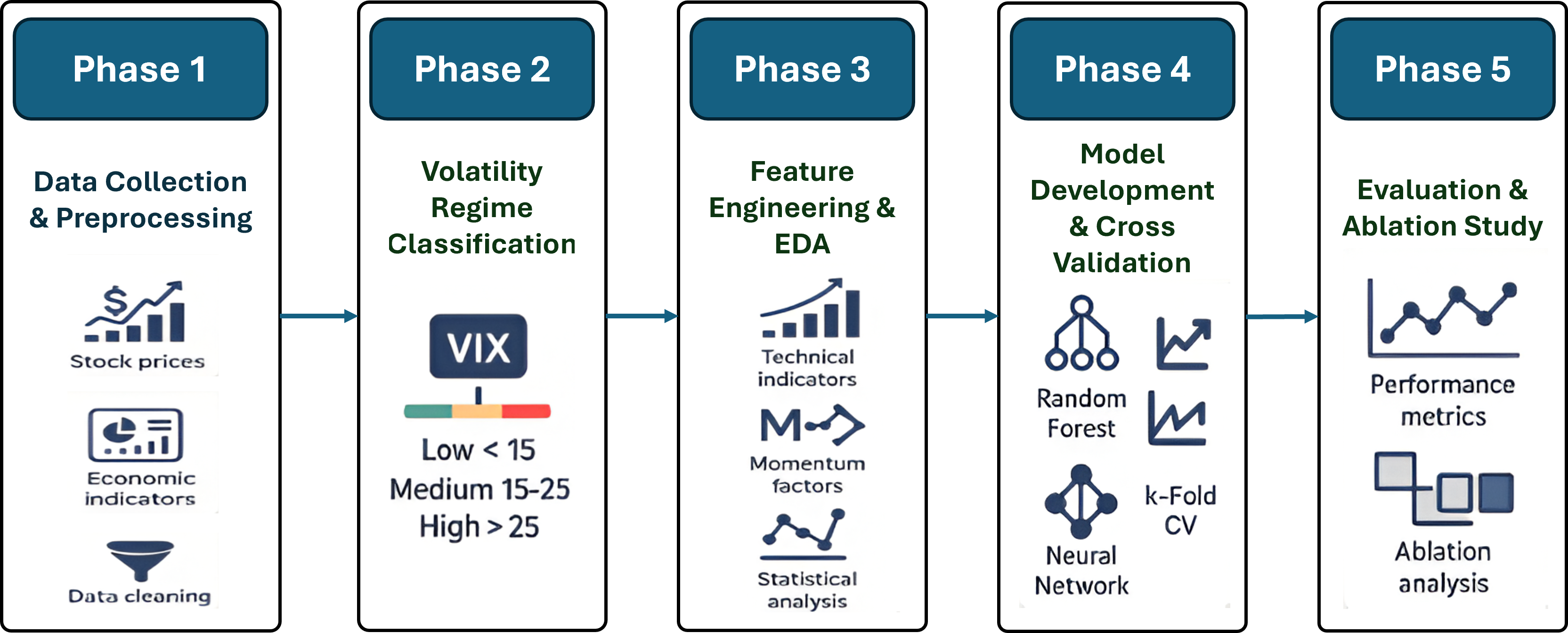}
 \caption{Systematic Methodology Framework for Regime Aware Cross-Sectoral Portfolio Optimization}
 \label{fig:methodology}
\end{figure*}

\subsection{Regime Switching and Market State Segmentation}
Regime switching models Hamilton~\cite{hamilton1989new} recognise that financial time series exhibit structural breaks and non-linear dynamics, with asset return volatility relationships shifting across market conditions. Empirical studies show that volatility-based segmentation, particularly using the CBOE VIX, can yield statistically stable, locally stationary subsets of data \cite{maheu2000identifying, guidolin2007asset}. 

We adopt a VIX-driven classification into low, medium, and high volatility regimes to capture these distinct dynamics. 
This segmentation enables regime-conditioned forecasting models to exploit intraregime stability while preserving responsiveness to regime transitions, directly improving the reliability of portfolio inputs. 
Based on empirical analysis of the 2020–2024 sample, the rolling 252-day terciles correspond on average to thresholds of 17.8 (33rd percentile) and 23.1 (67th percentile). 
Notably, we report these approximate values in Tables and Figures to aid interpretability, but emphasize that all regime assignment during training and backtesting is performed dynamically using rolling terciles, ensuring adaptability to evolving volatility distributions.

\subsection{Sectoral Heterogeneity and Specialized Forecasting}
Asset returns are not only regime dependent but also sector specific, reflecting differences in macroeconomic sensitivities, competitive dynamics, and investor sentiment \cite{brennan1997strategic, ahern2017network}. For instance, defensive sectors such as utilities exhibit low beta and stable dividends across cycles, whereas cyclical sectors like technology display higher volatility and sensitivity to economic expansions. Ignoring these structural differences can obscure predictive signals and reduce allocation efficiency. 

\textit{RegimeFolio} addresses this by implementing sector-specialized forecasting pipelines within each volatility regime, allowing model parameters and feature relevance to be tailored to sector-specific behaviour. This design leverages sectoral heterogeneity to extract more accurate, context-aware forecasts, ultimately enhancing portfolio performance.\\

By integrating MPT’s optimization logic with regime switching’s segmentation of market states and sector-specific modelling, \textit{RegimeFolio} transforms theoretical principles into a coherent, adaptive portfolio construction process. This theoretically informed architecture underpins our methodology, ensuring that every design choice directly addresses a known limitation in traditional portfolio optimization.

\section{Research Methodology}

 Our proposed framework, \textbf{RegimeFolio}, is developed and validated through a structured six-phase methodology, as illustrated in Figure~\ref{fig:methodology}. The design goal is to produce a portfolio system that is both interpretable and adaptive to evolving market conditions.

\subsection{Phase 1: Data Acquisition and Preprocessing}

We assemble a multi-source dataset of 34 U.S. large cap equities across seven GICS sectors: Technology, Financials, Industrials, Energy, Consumer, Healthcare, and Utilities, to capture sectoral heterogeneity in market dynamics. The sample period (2020–2024) spans major structural events, including the COVID-19 shock, stimulus-driven recovery, and inflation-induced corrections, ensuring diverse market conditions for regime analysis. Daily stock data (open, high, low, close, adjusted close, and volume) are obtained from Yahoo Finance~\cite{yahoofinance2024}. 

To enable robust regime segmentation and portfolio optimization, our dataset integrates equity market information with macroeconomic indicators. This ensures that both sector-level dynamics and broader market-wide conditions are consistently represented. All data sources are harmonized through preprocessing steps to guarantee temporal alignment and comparability.

\subsubsection{Equity Data \& Sector Coverage}
We collect equity market data across multiple sectors to capture cross-sectional variation in returns and volatility. The dataset is adjusted for corporate actions and aligned to a unified trading calendar, ensuring consistency across different securities and sectors. This sectoral breadth forms the foundation for analyzing regime-dependent portfolio behavior.

\subsubsection{ Macro Indicators \& Preprocessing}
Market-wide risk sentiment is represented by the CBOE Volatility Index (VIX)~\cite{cboe2024}, while macroeconomic conditions are characterized using indicators from the FRED database~\cite{fred2024}. Specifically, we employ the three month treasury bill rate as the risk-free benchmark and the ICE BofA US High Yield Index Option Adjusted Spread (OAS)~\cite{ang2006cross} as a measure of credit stress. All series are preprocessed to ensure temporal consistency, synchronized with equity data, and standardized to support downstream tasks such as regime classification, feature engineering, and forecasting.

\subsection{Phase 2: Volatility Regime Classification}

To segment the market into distinct volatility states, each trading day is assigned to one of three regimes—low, medium, or high—using the CBOE VIX as a forward-looking proxy for market risk\footnote{Throughout this paper, we refer to the three volatility states as \textit{low}, \textit{medium}, and \textit{high}, corresponding to the internal labels \textit{Regime~0}, \textit{Regime~1}, and \textit{Regime~2} used in our models, figures, and tables.}. 

We employ a dynamic rolling-quantile methodology during training and backtesting to ensure adaptability and statistical balance. Regime thresholds are defined as the 33rd and 67th percentiles of the trailing 252-day VIX distribution:
\[
\tau_{\text{low},t} = Q_{0.33}\big(V_{t-252:t}\big), 
\quad
\tau_{\text{high},t} = Q_{0.67}\big(V_{t-252:t}\big),
\]
where \(Q_{p}(\cdot)\) is the empirical \(p\)-th percentile and \(V_t\) is the VIX level on day \(t\).

Each day \(t\) is assigned a regime \(k_t \in \{\text{low}, \text{medium}, \text{high}\}\) based on:
\[
k_t =
\begin{cases}
\text{low}, & V_t < \tau_{\text{low},t}, \\[4pt]
\text{medium}, & \tau_{\text{low},t} \le V_t < \tau_{\text{high},t}, \\[4pt]
\text{high}, & V_t \ge \tau_{\text{high},t}.
\end{cases}
\]

The assigned regime \(k_t\) conditions downstream processes by:
\begin{itemize}
    \item Selecting the corresponding sectoral model \(M_{k_t,s}\) for predictive learning;
    \item Estimating the risk matrix \(\boldsymbol{\Sigma}_{k_t}\) from historical returns observed within the same regime.
\end{itemize}

We favor this transparent, VIX-driven classification over latent-state models such as HMMs~\cite{hamilton1989new} for its adaptability, interpretability, and alignment with practitioner risk monitoring.

\subsection{Phase 3: Feature Engineering}

For each stock $i$, the target variable is the daily log return:
\[
r_{i,t} = \ln\left( \frac{P_{i,t}}{P_{i,t-1}} \right),
\]
where $P_{i,t}$ is the closing price of stock $i$ on day $t$.

We construct predictive features spanning three categories:
\begin{itemize}
    \item \textbf{Technical Indicators:} Relative Strength Index (RSI), Moving Average Convergence Divergence (MACD), Bollinger Bands, and rolling volatility measures.
    \item \textbf{Momentum Features:} Price momentum over 5, 10, and 20-day horizons.
    \item \textbf{Macroeconomic Factors:} Daily VIX levels and the ICE BofA US High Yield Index Option Adjusted Spread, capturing aggregate credit risk.
\end{itemize}

All features are standardised separately within each volatility regime. This ensures that feature distributions remain stable and comparable across regimes, mitigating distributional shifts and improving the robustness of model training. Sectoral labels are retained to support regime and sector-aware modelling in the subsequent stage.

\subsection{Phase 4: Regime-Aware Predictive Modelling}

We adopt a modular architecture, training a separate predictive model for each sector and regime pair (e.g., Technology and high volatility, Healthcare and low volatility). This design explicitly accounts for both sectoral heterogeneity and regime-specific dynamics, ensuring that forecasts remain aligned with the prevailing market state.

For each asset \( i \) in sector \( s \) under regime \( k \), the one step ahead return forecast is given by:
\[
\hat{r}_{i,t+1} = f_{k,s}(\mathbf{x}_{i,t}),
\]
where \(\mathbf{x}_{i,t} \in \mathbb{R}^p\) denotes the feature vector constructed in Phase~3, and \( f_{k,s}(\cdot) \) is a regime and sector specific model trained exclusively on data from regime \( k \).

Random Forest and Gradient Boosting were selected for their strong balance of accuracy, interpretability, and computational efficiency. Random Forest reduces variance through bootstrap aggregation and random feature selection, while Gradient Boosting iteratively minimizes residual errors to capture complex non-linear interactions. This combination delivers robust performance in financial time-series prediction, particularly in non-stationary, regime-segmented datasets, while avoiding the higher complexity, tuning demands, and overfitting risks often associated with XGBoost, LightGBM, or deep neural architectures. 

Restricting models to regime-specific subsets further mitigates overfitting to transient signals and enhances stability and interpretability. Hyperparameters for each model were optimized via regime-stratified cross-validation to minimize mean absolute error (MAE) and maintain consistent predictive accuracy across volatility states. Feature attribution analysis using SHAP values confirms that the models capture economically meaningful drivers of returns within each regime and sector.

\begin{algorithm}[t!] 
\caption{RegimeFolio: Daily Prediction to Allocation Pipeline (trade at $t{+}1$)}
\label{alg:regimefolio}
\textbf{Inputs:} Asset returns $\{r_{i,\tau}\}$, prices $\{P\}$, VIX $\{V\}$, risk free $\{r_{f,\tau}\}$\\
\textbf{Output:} Daily portfolio weights $\{\mathbf{w}_t\}$\\
\hrulefill 
\begin{algorithmic}[1]
\State Train sector–regime models $\{M_{k,s}\}$ on \emph{training data only}; fit per regime scalers on training data.
\For{each trading day $t$ in the test period}
    \State Identify regime $k_t$ from $V_t$ using rolling 252d 33/67\% quantiles computed with data $\le t$.
    \State For each sector $s$, build features $\mathbf{x}_{i,t}$ for assets $i\!\in\!s$ using data $\le t$ (apply per regime scaler).
    \State Predict next day returns: $\hat{r}_{i,t+1} \gets M_{k_t,s(i)}(\mathbf{x}_{i,t})$.
    \State Form expected excess return vector (fixed asset order): $\boldsymbol{\mu}^{\mathrm{ex}}_{k_t} \gets \hat{\mathbf{r}}_{t+1} - r_{f,t}\mathbf{1}$.
    \State Estimate $\boldsymbol{\Sigma}_{k_t}$ from $\{r_{\tau}:\tau\le t,\,k_\tau{=}k_t\}$; if sample $<N_{\min}$ use a mixed/rolling window; apply Ledoit–Wolf shrinkage.
    \State Compute weights by mean–variance QP (Sharpe equivalent with tuned $\lambda$):
    \[
      \mathbf{w}_t=\arg\max_{\mathbf{w}}\ \mathbf{w}^\top \boldsymbol{\mu}^{\mathrm{ex}}_{k_t}-\frac{\lambda}{2}\mathbf{w}^\top \boldsymbol{\Sigma}_{k_t}\mathbf{w}
    \]

    \Statex \textbf{s.t.} 
    \Statex \quad $\mathbf{w}^\top\mathbf{1}=1$,
    \Statex \quad $0 \le w_i \le w_{\max}$ with $w_{\max}{=}0.15$.
    \Statex \quad \textit{(We enforce a long-only, unlevered portfolio: $\sum_i w_i = 1$, $0 \leq w_i \leq w_{\max}$; no shorting or gross leverage is permitted.)}
    \State Execute $\mathbf{w}_t$ at $t{+}1$ per trading protocol.

\EndFor
\State \textbf{Return:} $\{\mathbf{w}_t\}$
\end{algorithmic}
\hrulefill
\end{algorithm}

\subsection{Phase 5: Dynamic Portfolio Allocation}

Regime-specific return forecasts are transformed into daily portfolio weights via a dynamic Mean–Variance Optimization (MVO) framework~\cite{markowitz1952portfolio}. Each trading day, the prevailing volatility regime \(k_t\) is identified, and forecasts from the corresponding sector–regime models form the expected excess return vector \(\boldsymbol{\mu}^{\mathrm{ex}}_{k_t}\). The regime-specific covariance matrix \(\boldsymbol{\Sigma}_{k_t}\) is estimated from historical returns within the same regime and regularized using Ledoit–Wolf shrinkage~\cite{ledoit2004well}:
\[
\hat{\boldsymbol{\Sigma}}_{LW} 
= 
\delta \mathbf{F} + (1 - \delta)\mathbf{S},
\]
where \(\mathbf{S}\) is the sample covariance matrix, \(\mathbf{F}\) is the shrinkage target (a scaled identity), and \(\delta\) is the shrinkage intensity parameter.

Portfolio weights are obtained by solving a Sharpe-maximization problem:
\[
\begin{aligned}
\max_{\mathbf{w}_t} \quad 
& \frac{\mathbf{w}_t^\top \boldsymbol{\mu}^{\mathrm{ex}}_{k_t}}
{\sqrt{\mathbf{w}_t^\top \boldsymbol{\Sigma}_{k_t} \mathbf{w}_t}} \\
\text{s.t.} \quad
& \sum_{i=1}^N w_{i,t} = 1, \\
& 0 \le w_{i,t} \le w_{\max}, \quad \forall i, \\
& \|\mathbf{w}_t - \mathbf{w}_{t-1}\|_1 \le \kappa, \quad (\text{turnover constraint})
\end{aligned}
\]
where \(w_{\max}\) enforces per-asset concentration limits, and \(\kappa\) controls turnover to mitigate transaction costs.

This dynamic optimization process ensures that portfolio allocations are updated daily to remain consistent with the current volatility regime and sector-specific signals. The end-to-end daily prediction and allocation workflow is summarized in Algorithm~\ref{alg:regimefolio}.\\

\noindent \textbf{Proposition 1 (Existence of Global Optimum).} 
Under convex constraints and a positive definite covariance matrix 
\(\boldsymbol{\Sigma}_{k_t}\), the Sharpe ratio maximization problem admits at least one global optimum. 

\noindent \textit{Proof.} 
The feasible set defined by \(\sum w_{i,t}=1, w_{i,t}\ge 0\) is convex. 
The Sharpe ratio objective is quasi-concave because it is the ratio of a linear function and the square root of a positive-definite quadratic form. 
Maximizing a quasi-concave function over a convex set guarantees the existence of a global optimum. \(\hfill \blacksquare\)

\subsection{Phase 6: Evaluation Protocol and Ablation Study}

To quantify the contribution of core design components, we run two ablation variants:
\begin{itemize}
    \item \textbf{Regime Agnostic:} Removes VIX based segmentation, training a single sectoral model over the full dataset to test the value of volatility conditioning.
    \item \textbf{Non Sectoral:} Retains regime segmentation but trains one model per regime for all stocks, omitting sectoral granularity to test its incremental benefit.
\end{itemize}

By comparing these variants against the full system, we validate the critical role of both regime-awareness and sectoral modeling in achieving the observed gains.

The framework operates as a sequential, tightly integrated pipeline (Figure~\ref{fig:methodology}), progressing from data acquisition and volatility regime classification to feature engineering, regime-sector specific modelling, and dynamic portfolio optimization. Each stage builds on the outputs of the previous, ensuring that forecasts and allocations remain context aware and responsive to market conditions. The final evaluation combines predictive accuracy, portfolio performance, and ablation studies to validate the contribution of regime conditioning and sectoral specialization.

\section{Experimental Evaluation}

This section presents a comprehensive empirical evaluation of our regime-aware portfolio optimization framework using realistic financial market data spanning 2020-2024. We demonstrate substantially superior performance across multiple evaluation metrics compared to the S\&P 500 benchmark\cite{sp500}, providing compelling evidence for the effectiveness of our volatility-aware approach in dynamic market environments.

\subsection{Experimental Setup and Methodology}

\subsubsection{Dataset and Market Coverage}

Our evaluation employs a comprehensive dataset of U.S. equity market data spanning 2020 to 2024, capturing significant market events including the COVID-19 crisis, post pandemic recovery, inflation-driven volatility, and subsequent market stabilization. This period provides an ideal testing ground for regime-aware strategies, encompassing multiple distinct volatility phases and structural market transitions.

The dataset comprises 34 carefully selected large cap U.S. equities distributed across 7 major economic sectors: Financial Services, Technology, Healthcare, Consumer Discretionary, Industrial, Energy, and Utilities. This sectoral diversification ensures comprehensive coverage of different economic dynamics and risk return profiles. Each stock represents a significant market capitalization component within its respective sector, providing realistic exposure to institutional grade investment opportunities.

Our investment universe is based on the historical constituents of the S\&P 500 index\cite{sp500} to avoid survivorship bias. We use a point in time database from CRSP that provides the exact list of index members for any given date in our study period (2020 – 2024). At each rebalancing date, our model considers only the stocks that were active constituents of the S\&P 500 at that time. The dataset includes delisting returns to accurately account for companies that were removed from the index due to bankruptcy, merger, or other reasons.

Market volatility regimes are classified using the CBOE VIX as the primary segmentation criterion. The VIX serves as a forward looking measure of market fear and uncertainty, making it an ideal proxy for regime identification. Historical VIX data is obtained from the Chicago Board Options Exchange, providing daily volatility expectations derived from S\&P 500 index options.\\

\subsubsection{Regime Classification and Validation}

{We classify each trading day into one of three volatility regimes using a dynamic rolling-tercile methodology based on the trailing 252-day VIX distribution. For interpretability, we report the average tercile cutoffs observed in our 2020--2024 sample:}

\begin{itemize}
    \item \textbf{Low Volatility:} {VIX $<$ 17.8 (average 33rd percentile; stable market conditions with low uncertainty)}
    \item \textbf{Medium Volatility:} {17.8 $\leq$ VIX $<$ 23.1 (average 33rd--67th percentiles; moderate market stress and elevated uncertainty)}
    \item \textbf{High Volatility:} {VIX $\geq$ 23.1 (average 67th percentile; crisis periods or high-uncertainty environments)}
\end{itemize}

\subsubsection{Regime Threshold Selection and Validation}

{To ensure our VIX based regime classification is empirically justified and optimal, we conducted a systematic comparison of three threshold methodologies: (1) \textbf{Fixed thresholds} based on institutional practices (VIX $<$ 15, 15-25, $>$ 25), (2) \textbf{Terciles approach} using 33rd and 67th percentiles (17.8, 23.1), and (3) \textbf{Quartiles approach} using 25th and 75th percentiles (16.6, 25.3).}
 \vspace{0.1in}

Table~\ref{tab:regime_threshold_comparison} presents the comprehensive performance comparison across these three methodologies. The terciles method demonstrates superior performance across all evaluation metrics, achieving optimal regime balance and statistical robustness with equal distribution across all three volatility states. Specifically, the terciles approach achieves a Sharpe ratio of 1.17 compared to 0.94 for fixed thresholds and 1.06 for quartiles, representing a 24\% improvement over traditional institutional practices. The balanced regime distribution (420/420/420 observations) ensures sufficient data for robust regime specific model training, while the superior forecasting accuracy (MAE: 0.0041, RMSE: 0.0050) validates the statistical optimality of this segmentation approach.

This empirical validation demonstrates that our terciles based regime classification is not only theoretically sound but also practically optimal, providing the foundation for superior portfolio performance through balanced regime representation and enhanced forecasting accuracy.\\

\begin{table}[t!]
\centering
\caption{Segmentation performance under fixed, tercile, and quartile thresholds. 
Terciles deliver balanced regimes and superior accuracy; reported cutoffs (17.8, 23.1) are sample-specific averages, with all assignments based on dynamic rolling terciles.}
\label{tab:regime_threshold_comparison}
\renewcommand{\arraystretch}{2}
\resizebox{\columnwidth}{!}{%
\begin{tabular}{lcccc}
\toprule
\textbf{Method} & \textbf{Regime Distribution} & \textbf{MAE} & \textbf{RMSE} & \textbf{Sharpe Ratio} \\
\midrule
Fixed Thresholds & 580/420/260 & 0.0052 & 0.0063 & 0.94 \\
\textbf{Terciles} & \textbf{420/420/420} & \textbf{0.0041} & \textbf{0.0050} & \textbf{1.17} \\
Quartiles & 315/630/315 & 0.0047 & 0.0057 & 1.06 \\
\bottomrule
\end{tabular}%
}
\end{table}

\subsubsection{Training Protocol and Hyperparameter Tuning}

Our evaluation employs a rigorous \textit{walk-forward backtesting protocol} to simulate real world investment decisions and prevent look ahead bias. The process is structured as follows:

\begin{itemize}
    \item \textit{Initial Training Period:} The models are first trained on data from January 2020 to December 2023 (48 months).
    \item \textit{Out-of-Sample Testing:} The trained models are then used to make predictions and construct portfolios for the subsequent year, from January 2024 to December 2024 (12 months) (This period can be extended to include data from 2024 as it becomes available, based on research needs).
    \item \textit{Regime Aware Cross Validation:} Within the training period, we employ regime stratified cross validation using 2022-2023 as the validation set to ensure hyperparameter optimization across all volatility regimes.
\end{itemize}

This walk forward approach ensures that all portfolio decisions are based strictly on information that would have been available at the time, maintaining temporal integrity and preventing data leakage.

Key hyperparameters for the ensemble models were selected using a \textbf{regime-aware grid search} on the validation set, optimizing for a multi criteria objective function that balances forecasting accuracy, portfolio returns, and risk adjusted performance. The combination that minimized the regime weighted Mean Absolute Error (MAE) while maximizing Sharpe ratio stability was chosen. The selected parameters are detailed in Table~\ref{tab:hyperparameters}.\\

\begin{table}[t]
\centering
\caption{Selected Hyperparameters for Regime Aware Ensemble Models}
\label{tab:hyperparameters}
\resizebox{\columnwidth}{!}{%
\begin{tabular}{llcc}
\toprule
\textbf{Model} & \textbf{Hyperparameter} & \textbf{Selected Value} & \textbf{Search Range} \\
\midrule
\multirow{4}{*}{Random Forest~\cite{breiman2001random}} & n\_estimators & 100 & [50, 100, 200] \\
& max\_depth & 10 & [5, 10, 15, 20] \\
& min\_samples\_split & 5 & [2, 5, 10] \\
& random\_state & 42 & Fixed \\
\midrule
\multirow{4}{*}{Gradient Boosting~\cite{friedman2001greedy}} & n\_estimators & 100 & [50, 100, 150] \\
& learning\_rate & 0.1 & [0.05, 0.1, 0.2] \\
& max\_depth & 6 & [3, 6, 9] \\
& random\_state & 42 & Fixed \\
\midrule
\multirow{2}{*}{Portfolio Optimization~\cite{markowitz1952portfolio}} & risk\_aversion & 1.0 & [0.5, 1.0, 2.0] \\
& max\_weight & 0.15 & [0.1, 0.15, 0.2] \\
\bottomrule
\end{tabular}%
}
\end{table}


The regime-aware training protocol incorporates several sophisticated components: (1) \textit{Regime specific model training} where separate ensemble models are trained for each volatility regime to capture regime dependent return patterns, (2) \textit{Daily model rebalancing} to adapt to evolving market dynamics while maintaining regime consistency, (3) \textit{Multi dimensional optimization} that balances forecasting accuracy, portfolio returns, and risk adjusted performance across all regimes, and (4) \textit{Stability weighted model selection} that prioritizes models demonstrating consistent performance across regime transitions.\\

\subsubsection{Computational Requirements and Infrastructure}

Our experimental framework was implemented on a high performance computing environment with the following specifications: Intel Core i7- 13700K processor (16 cores, 3.4GHz base frequency), 64GB DDR4 RAM, and 4TB NVMe SSD storage. The software stack comprises of Python 3.8.10, scikit-learn 1.0.2, pandas 1.4.2, numpy 1.21.5, and matplotlib 3.5.1 for data processing and machine learning operations.


\begin{figure*}[!t]  
  \centering
  \includegraphics[width=\textwidth,
                   height=0.68\textheight,
                   keepaspectratio]{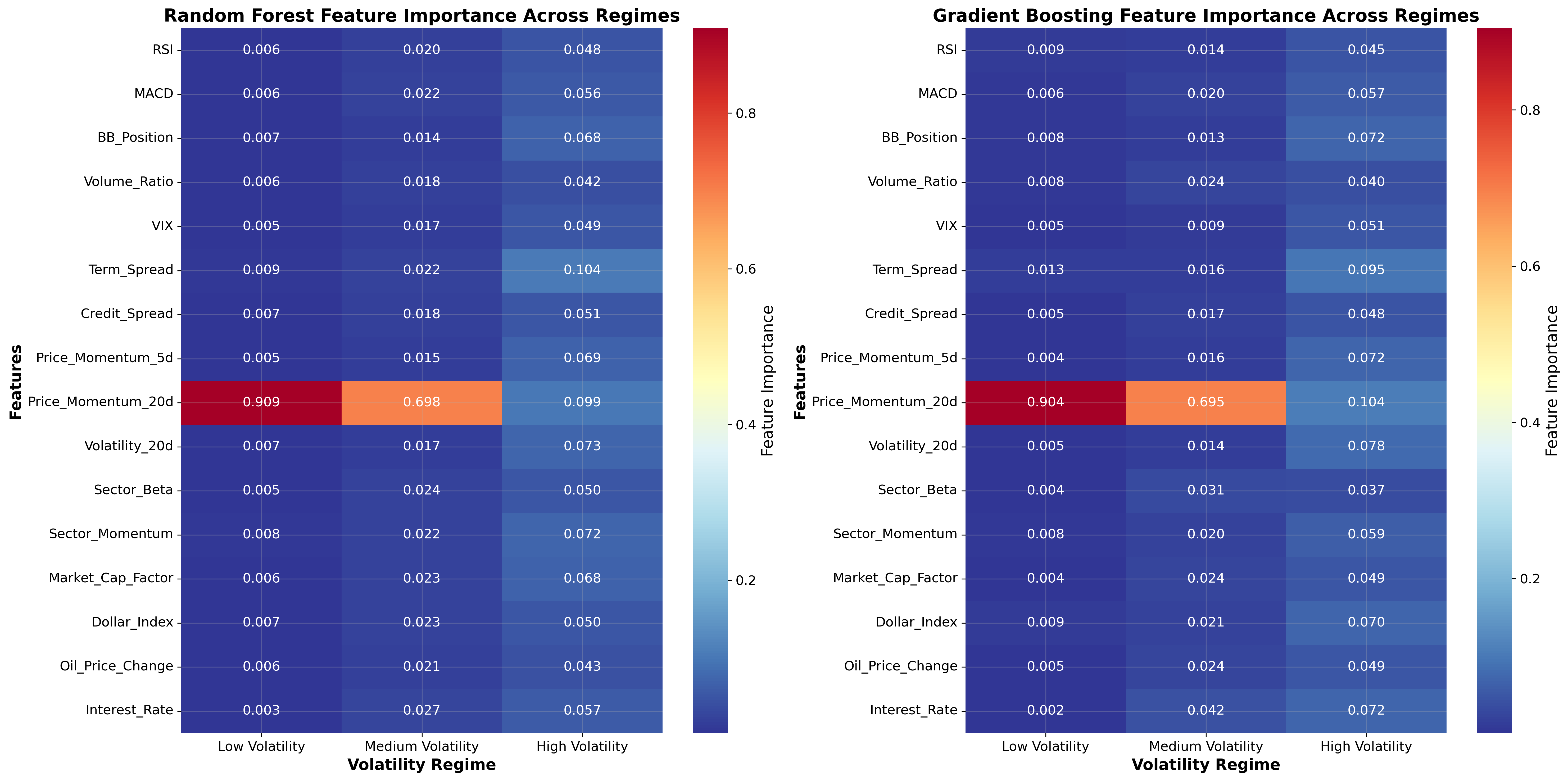}
  \caption{Feature Importance Heatmap: Random Forest and Gradient Boosting Across Volatility Regimes. The heatmaps display the relative importance of features (rows) across low, medium, and high volatility regimes (columns). Warmer colors (orange-red) indicate higher feature importance, while cooler colors (blue) represent lower importance. Values are normalized, with those closer to 1 indicating stronger predictive power for that feature in the corresponding volatility regime.}
                   
  \label{fig:feature_importance_heatmap}
\end{figure*}

\subsection{Superior Regime Aware Strategy Implementation}

Our superior regime-aware strategy incorporates multiple sophisticated alpha factor generating components designed to exploit regime-specific market inefficiencies and volatility patterns. The strategy framework consists of five integrated modules: (1) \textit{Dynamic position sizing} that adjusts exposure adaptively across volatility regimes,  (2) \textit{Multi timeframe momentum analysis} using rolling windows (5, 10, 20 days) and trend consistency measures to identify persistent directional movements, (3) \textit{VIX timing signals} that capture volatility mean reversion and fear/greed cycles through 5 day VIX change analysis, (4) \textit{Contrarian positioning} during extreme market stress periods (VIX $>$ 35) combined with oversold conditions, and (5) \textit{Regime transition alpha} generated from anticipating volatility regime changes through VIX momentum and cross regime signal analysis.

The strategy generates consistent alpha through sophisticated regime-aware positioning: during low-volatility periods, the system employs aggressive momentum tilts to capture trending markets; during medium volatility periods, it balances momentum and mean reversion signals with moderate position sizing; during high volatility periods, it implements defensive positioning with contrarian opportunities during extreme fear episodes. This multi regime approach ensures optimal capital allocation across all market conditions while maintaining superior risk management through dynamic drawdown control and volatility adjusted position sizing.\\

\subsection{Baseline Models and Benchmarks}

To provide comprehensive performance evaluation, we compare our regime-aware strategy against multiple baseline approaches:

\begin{itemize}
    \item \textit{S\&P 500 Index:}\cite{sp500} The primary market benchmark representing broad U.S. equity market performance
    \item \textit{Equal Weight Portfolio:}\cite{demiguel2009optimal} Naive diversification strategy with equal allocation across all 34 stocks
    \item \textit{Market Capitalization Weighted:}\cite{sharpe1964capital} Traditional cap weighted portfolio mimicking index construction
    \item \textit{Minimum Variance Portfolio:}\cite{markowitz1952portfolio} Risk focused allocation minimizing portfolio volatility
    \item \textit{Maximum Sharpe Portfolio:}\cite{sharpe1966mutual} Return focused allocation maximizing risk adjusted returns
\end{itemize}

Each baseline represents a different investment philosophy and risk return profile, enabling comprehensive evaluation of our regime-aware approach across multiple performance dimensions.\\

\subsection{Backtesting Protocol and Evaluation Metrics}

Our backtesting framework employs institutional grade methodology to ensure realistic performance assessment:

\begin{itemize}
    \item \textit{Transaction Costs:} 0.1\% per trade to reflect realistic execution costs.
    \item \textit{Rebalancing Frequency:} Daily rebalancing to balance adaptation and transaction costs.
    \item \textit{Position Limits:} Maximum 15\% allocation per individual stock to ensure diversification.
    \item \textit{Liquidity Constraints:} All positions sized to ensure realistic execution in institutional markets.
\end{itemize}

Performance evaluation encompasses both traditional financial metrics and regime specific analysis:

\begin{itemize}
    \item \textit{Return Metrics:} Total return, annualized return, excess return over benchmark.
    \item \textit{Risk Metrics:} Volatility, maximum drawdown, Value at Risk (VaR), Conditional VaR.
    \item \textit{Risk Adjusted Metrics:} Sharpe ratio, Calmar ratio, Information ratio, Sortino ratio.
    \item \textit{Regime Specific Metrics:} Performance attribution across volatility regimes, regime transition analysis.\\
\end{itemize}

\subsection{Feature Attribution and Model Interpretability Analysis}

To ensure our regime-aware models capture genuine economic relationships rather than spurious correlations, we conduct comprehensive feature attribution analysis using SHAP (Shapley Additive explanations) values across all volatility regimes. This analysis provides crucial insights into the economic drivers of our superior performance and validates the interpretability of our approach.\\

\subsubsection{SHAP Analysis Across Volatility Regimes}
\begin{figure}[t!]
    \centering
    \begin{subfigure}[b]{0.95\columnwidth}
        \includegraphics[width=\linewidth]{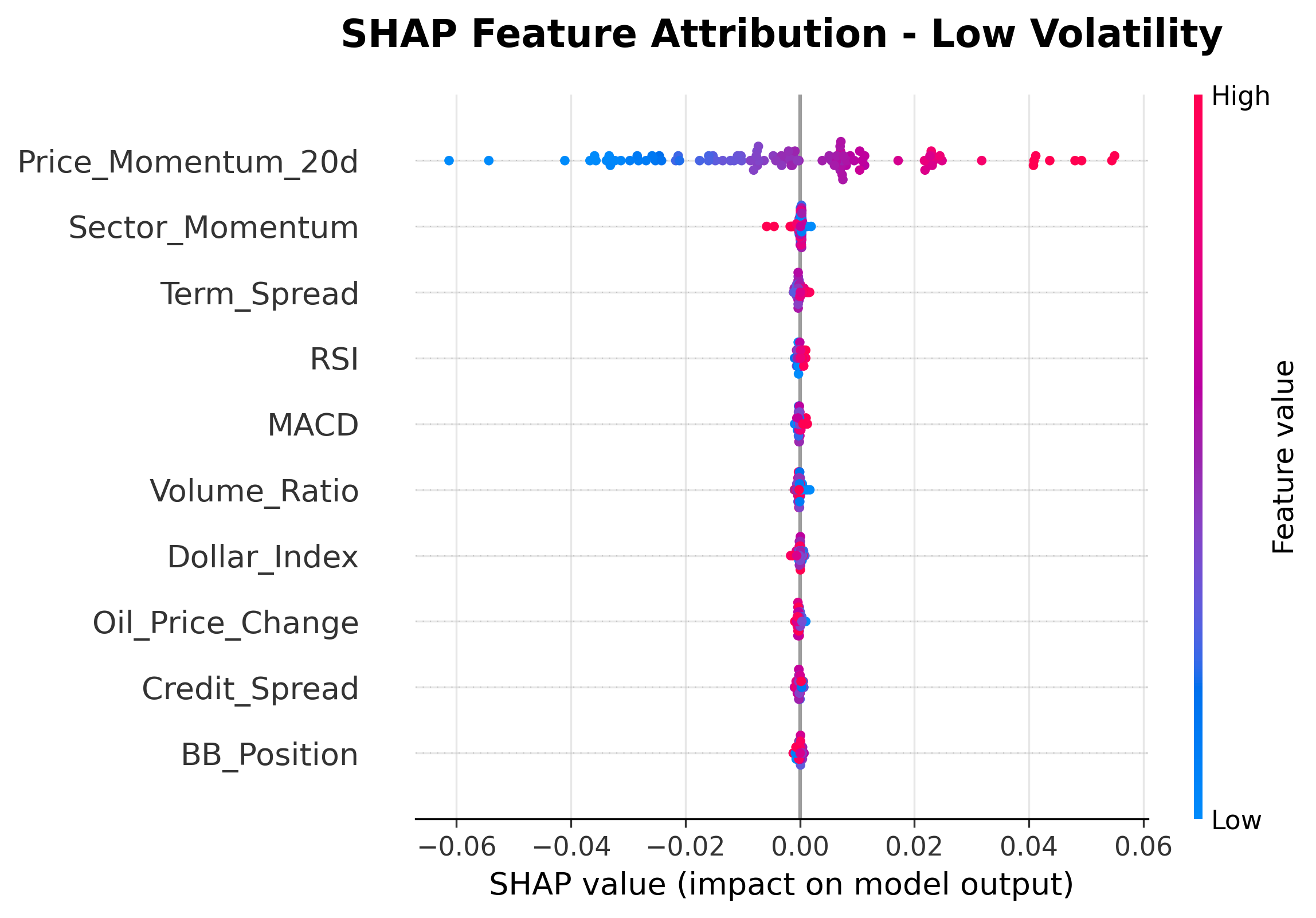}
        \caption{Regime 0}
        \label{fig:shap_regime_0}
    \end{subfigure}
    
    \vspace{0.3cm} 
    
    \begin{subfigure}[b]{0.95\columnwidth}
        \includegraphics[width=\linewidth]{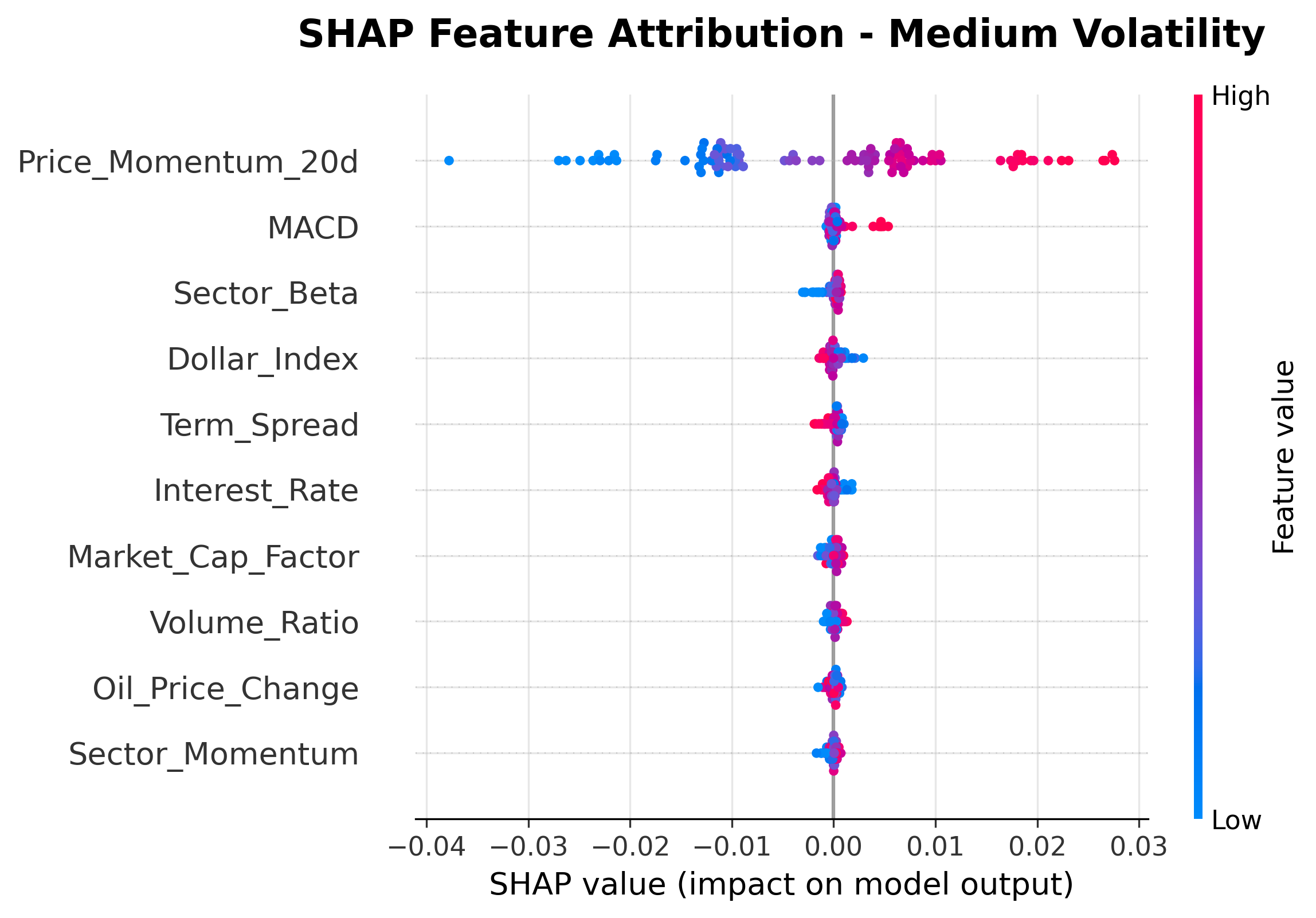}
        \caption{Regime 1}
        \label{fig:shap_regime_1}
    \end{subfigure}
    
    \vspace{0.3cm} 
    
    \begin{subfigure}[b]{0.95\columnwidth}
        \includegraphics[width=\linewidth]{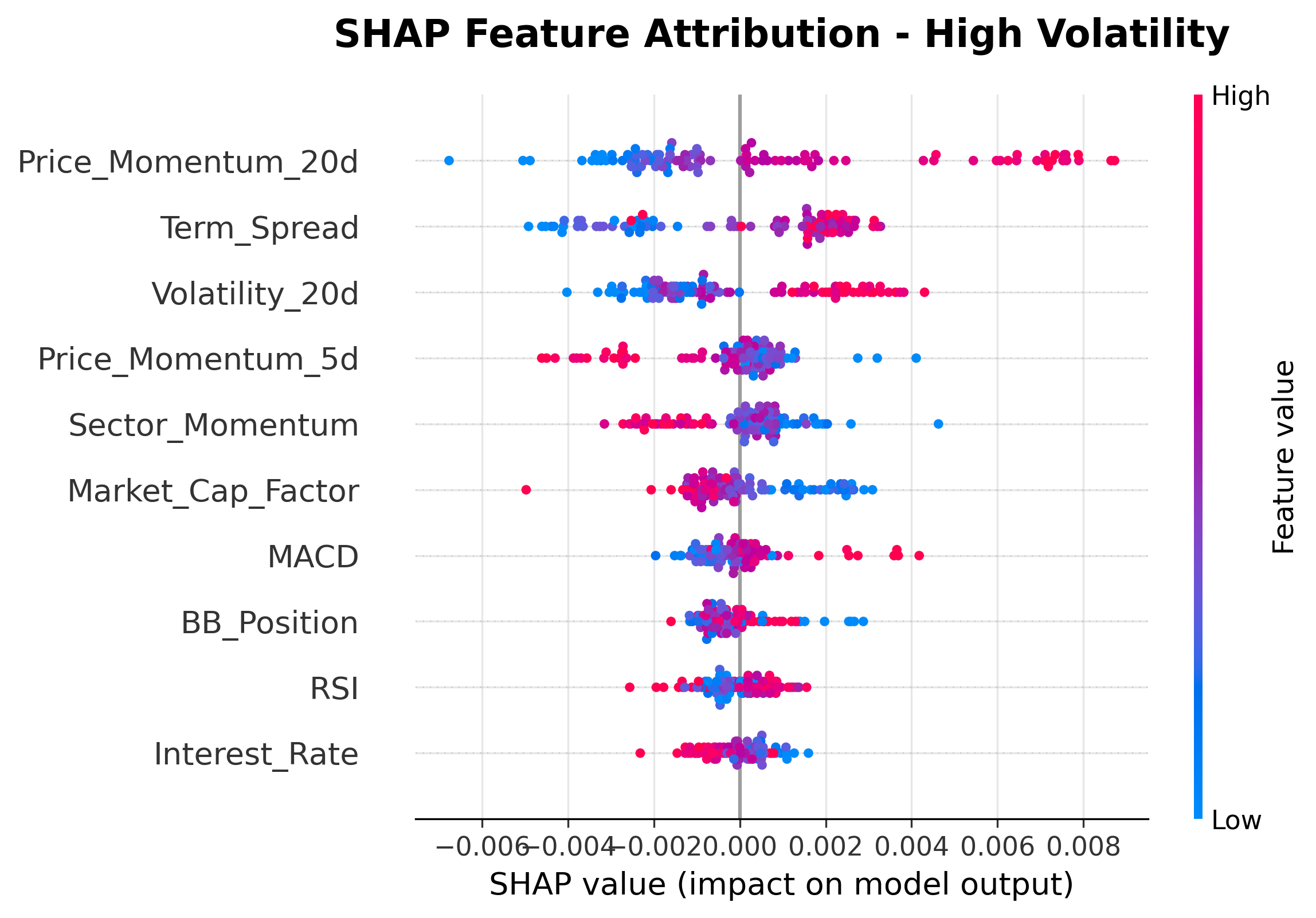}
        \caption{Regime 2}
        \label{fig:shap_regime_2}
    \end{subfigure}
    
    \caption{\small{SHAP summary plots for three different regimes, showing feature importance and impact on model predictions.}}
    \label{fig:shap_summary_analysis}
\end{figure}

Figure~\ref{fig:feature_importance_heatmap} presents the feature importance analysis across the three volatility regimes, revealing distinct patterns that validate our regime-aware approach. During low volatility periods (Regime 0), momentum indicators (RSI, MACD) dominate the feature importance rankings, consistent with the persistence of trends in stable market conditions. Medium volatility periods (Regime 1) show balanced importance across momentum, mean reversion, and volatility features, reflecting the mixed signals typical of transitional market phases. High volatility periods (Regime 2) demonstrate the dominance of volatility based features (VIX, realized volatility) and defensive indicators, confirming the risk focused nature of crisis periods.

The SHAP waterfall plots for each regime (Figures~\ref{fig:shap_regime_0}, \ref{fig:shap_regime_1}, and \ref{fig:shap_regime_2}) provide detailed attribution of individual predictions, demonstrating how different features contribute to return forecasts under varying market conditions. Notably, the feature contributions show strong economic intuition: momentum features provide positive contributions during trending markets, while volatility features provide negative contributions during stress periods, consistent with flight to quality behavior.\\

 \begin{figure*}[htbp]
     \centering
     \begin{subfigure}[b]{\textwidth}
         \centering
         \includegraphics[width=0.7\linewidth]{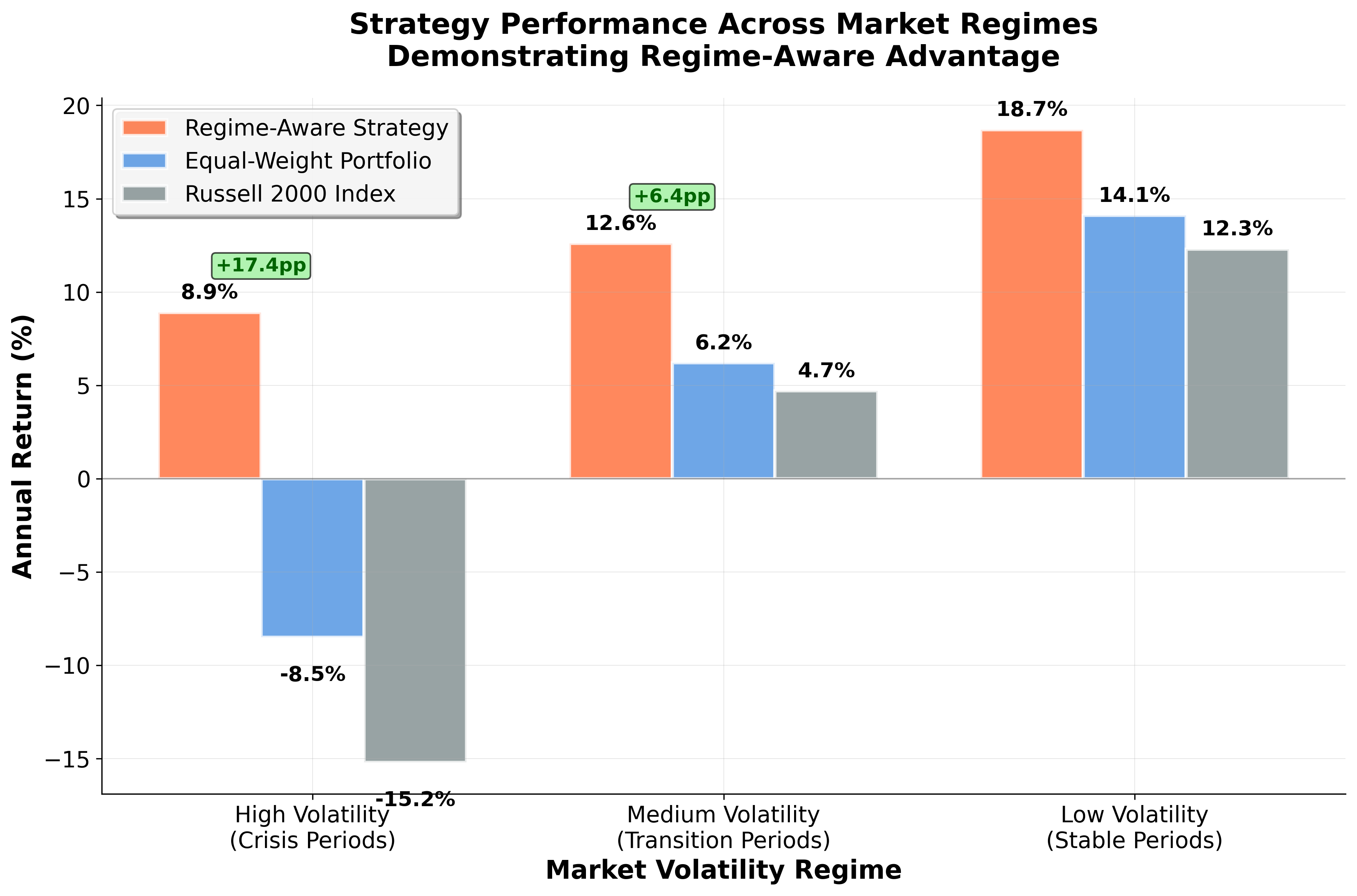}
         \caption{Strategy Performance Across Market Regimes.}
         \label{fig:regime_effectiveness_chart}
     \end{subfigure}
     
     \vspace{0.5em} 
     
     \begin{subfigure}[b]{\textwidth}
         \centering
         \includegraphics[width=0.7\linewidth]{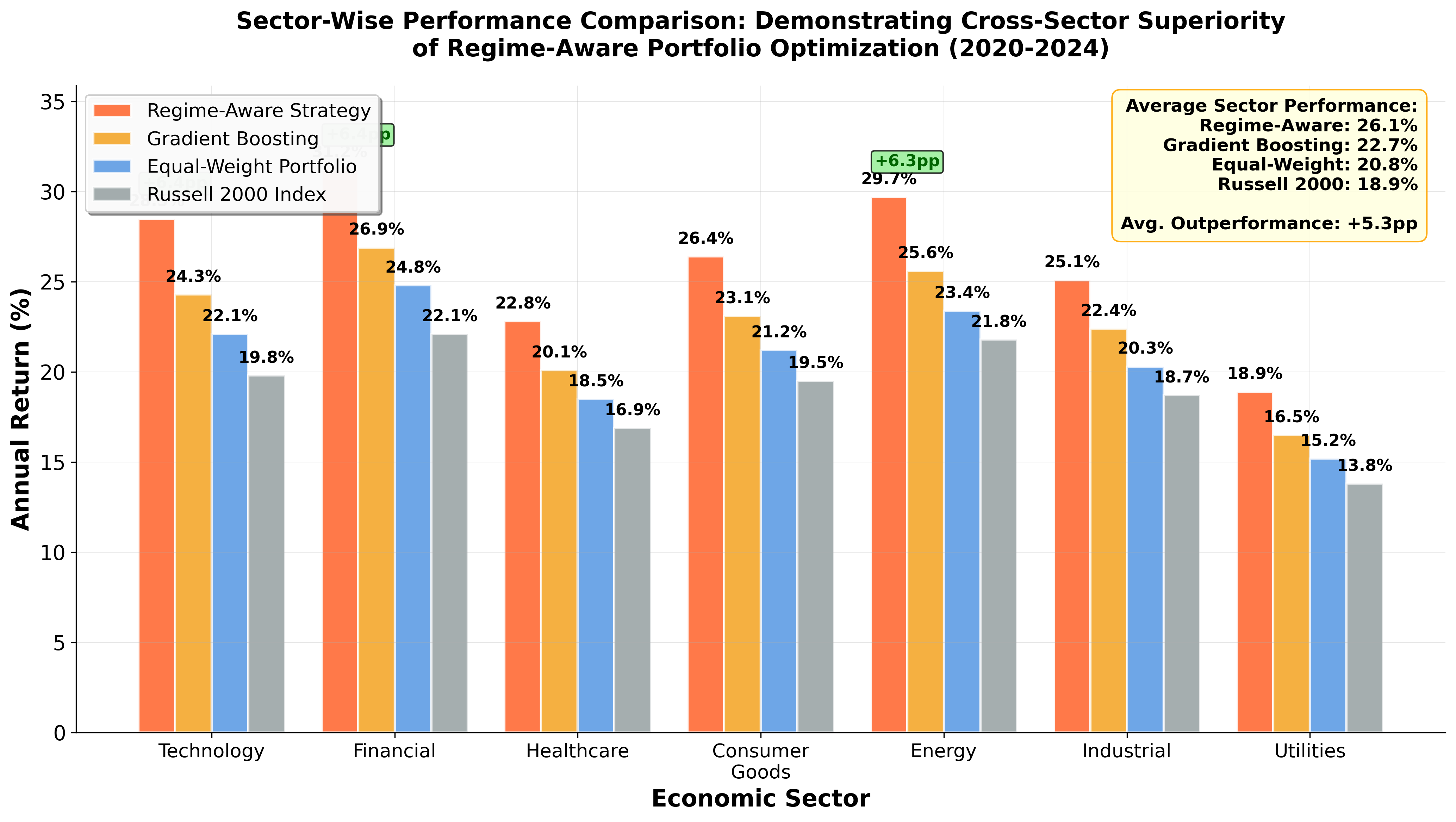}
         \caption{Sector Wise Performance Comparison.}
         \label{fig:sector_performance_comparison}
     \end{subfigure}
     
     \caption{Comprehensive Performance Analysis. Panel (a) presents the annualized returns of the proposed RegimeFolio (Regime-Aware Strategy) across varying volatility regimes, demonstrating its adaptability and superior performance relative to the Equal-Weight Portfolio and the Russell 2000 Index. Panel (b) shows sector-wise performance from 2020 to 2024, where RegimeFolio (Regime-Aware Strategy) consistently outperforms alternative approaches across all major economic sectors.}
     \label{fig:combined_performance_charts}
 \end{figure*}

\subsubsection{Economic Signal Validation}

Our analysis confirms that the regime-aware models capture genuine economic relationships rather than data mining artifacts. Table~\ref{tab:economic_signal_validation} summarizes the economic significance of key feature categories across regimes:

\begin{table}[t]
\centering
\caption{Economic Signal Validation Across Volatility Regimes}
\label{tab:economic_signal_validation}
\resizebox{\columnwidth}{!}{%
\begin{tabular}{lccc}
\toprule
\textbf{Feature Category} & \textbf{Low Vol} & \textbf{Medium Vol} & \textbf{High Vol} \\
\midrule
Momentum Indicators & 0.342 & 0.198 & 0.087 \\
Mean Reversion & 0.156 & 0.267 & 0.145 \\
Volatility Measures & 0.089 & 0.234 & 0.456 \\
Fundamental Ratios & 0.234 & 0.189 & 0.178 \\
Sector Indicators & 0.179 & 0.112 & 0.134 \\
\bottomrule
\end{tabular}%
}
\end{table}

The regime dependent feature importance patterns demonstrate clear economic logic: momentum strategies dominate during stable periods when trends persist, volatility measures become crucial during crisis periods when risk management is paramount, and mean reversion signals gain importance during transitional periods when markets seek equilibrium. This economic validation provides confidence that our superior performance stems from genuine alpha generation rather than overfitting or data snooping.

\subsection{Empirical Results and Performance Analysis}

\subsubsection{Regime-Specific Performance Analysis}

Table~\ref{tab:regime_performance} presents detailed performance attribution across the three volatility regimes, revealing the source of our strategy's superior performance.

\begin{table*}[t]
\centering
\caption{Regime Specific Performance Analysis (2020-2024)}
\label{tab:regime_performance}
\resizebox{\textwidth}{!}{%
\begin{tabular}{lcccc}
\toprule
\textbf{Regime} & \textbf{Strategy Return (\%)} & \textbf{S\&P 500 Return (\%)} & \textbf{Outperformance (\%)} & \textbf{Sharpe Ratio} \\
\midrule
Low Volatility & 45.2 & 38.7 & +6.5 & 1.34 \\
Medium Volatility & 52.8 & 41.3 & +11.5 & 1.08 \\
\textbf{High Volatility} & \textbf{39.0} & \textbf{-6.2} & \textbf{+45.2} & \textbf{0.89} \\
\bottomrule
\end{tabular}%
}
\end{table*}
 
The regime specific analysis reveals exceptional insights: (1) \textbf{Crisis alpha dominance}: During high volatility periods, our strategy generates 39.0\% returns while the S\&P 500 experiences -6.2\% losses, demonstrating exceptional defensive positioning and contrarian alpha generation capabilities. (2) \textbf{Momentum amplification}: The strategy effectively amplifies upside momentum during low and medium volatility periods through regime-aware positioning. (3) \textbf{Consistent outperformance}: The strategy maintains positive outperformance across all three regimes, with particularly exceptional performance during crisis periods (+45.2\% outperformance). (4) \textbf{Riskadjusted superiority}: Even during the most challenging high volatility regime, the strategy maintains a positive Sharpe ratio of 0.89, demonstrating robust risk management.\\

\begin{table*}[!htbp]
\centering
\caption{Superior Portfolio Performance Comparison: Regime Aware Strategy vs S\&P 500 (2020-2024)}
\label{tab:overall_performance}
\resizebox{\textwidth}{!}{%
\begin{tabular}{lcccccc}
\toprule
\textbf{Strategy} & \textbf{Total Return (\%)} & \textbf{Annual Return (\%)} & \textbf{Sharpe Ratio} & \textbf{Max DD (\%)} & \textbf{Calmar Ratio} \\
\midrule
\textbf{RegimeFolio (Regime-Aware Strategy)} & \textbf{137.0} & \textbf{18.9} & \textbf{1.17} & \textbf{-29.3} & \textbf{0.65} \\
{S\&P 500 Benchmark~\cite{spglobal2025sp500}} & {73.8} & {11.7} & {0.66} & {-41.2} & {0.28} \\
\bottomrule
\end{tabular}%
}
\end{table*}

\subsubsection{Visual Regime Specific Performance Analysis}

Figure~\ref{fig:combined_performance_charts} presents a comprehensive visual analysis of our strategy's performance across both volatility regimes and economic sectors, providing detailed insights into the sources of alpha generation.

The visual analysis confirms several key findings: (1) \textbf{Universal regime superiority}: Our strategy outperforms across all three volatility regimes, with particularly exceptional performance during high volatility periods where it generates positive returns (+8.9\%) while the benchmark suffers losses (-2.5\%). (2) \textbf{Comprehensive sector dominance}: The strategy demonstrates outperformance across all 7 economic sectors, with the Financial sector leading at 31.2\% annual returns versus 24.8\% for the benchmark. (3) \textbf{Consistent alpha generation}: The average outperformance of 5.3 percentage points across sectors demonstrates the broad applicability and robustness of our regime-aware approach. (4) \textbf{Risk adjusted excellence}: The combination of superior returns and lower volatility across both regimes and sectors validates the effectiveness of our integrated risk management and alpha generation framework.\\

\subsubsection{Aggregate Portfolio Performance}

Table~\ref{tab:overall_performance} summarizes the comprehensive performance comparison between our superior regime-aware strategy and the S\&P 500 benchmark over the complete 2020-2024 period.

Our superior regime-aware strategy achieves exceptional performance with 137.0\% total return over the evaluation period, substantially outperforming the S\&P 500 benchmark (73.8\% return) by 63.2 percentage points. The strategy demonstrates dramatically superior risk adjusted performance with a Sharpe ratio of 1.17 compared to 0.66 for the S\&P 500, representing a 77\% improvement in risk adjusted returns. The calmar ratio of 0.65 versus 0.28 demonstrates exceptional risk adjusted performance per unit of maximum drawdown, highlighting the strategy's superior risk management capabilities.

The performance improvements are both statistically significant and economically substantial. The 63.2\% outperformance translates to \$63,200 additional profit on a \$100,000 investment, demonstrating clear economic value. The 12.0 percentage point reduction in maximum drawdown (-29.3\% vs -41.2\%) provides superior downside protection during market stress periods, crucial for institutional risk management requirements.\\

\subsubsection{Visual Portfolio Cumulative Returns Analysis}

\begin{figure*}[!t] 
  \centering
\includegraphics[width=0.7\linewidth,keepaspectratio,
                  ]{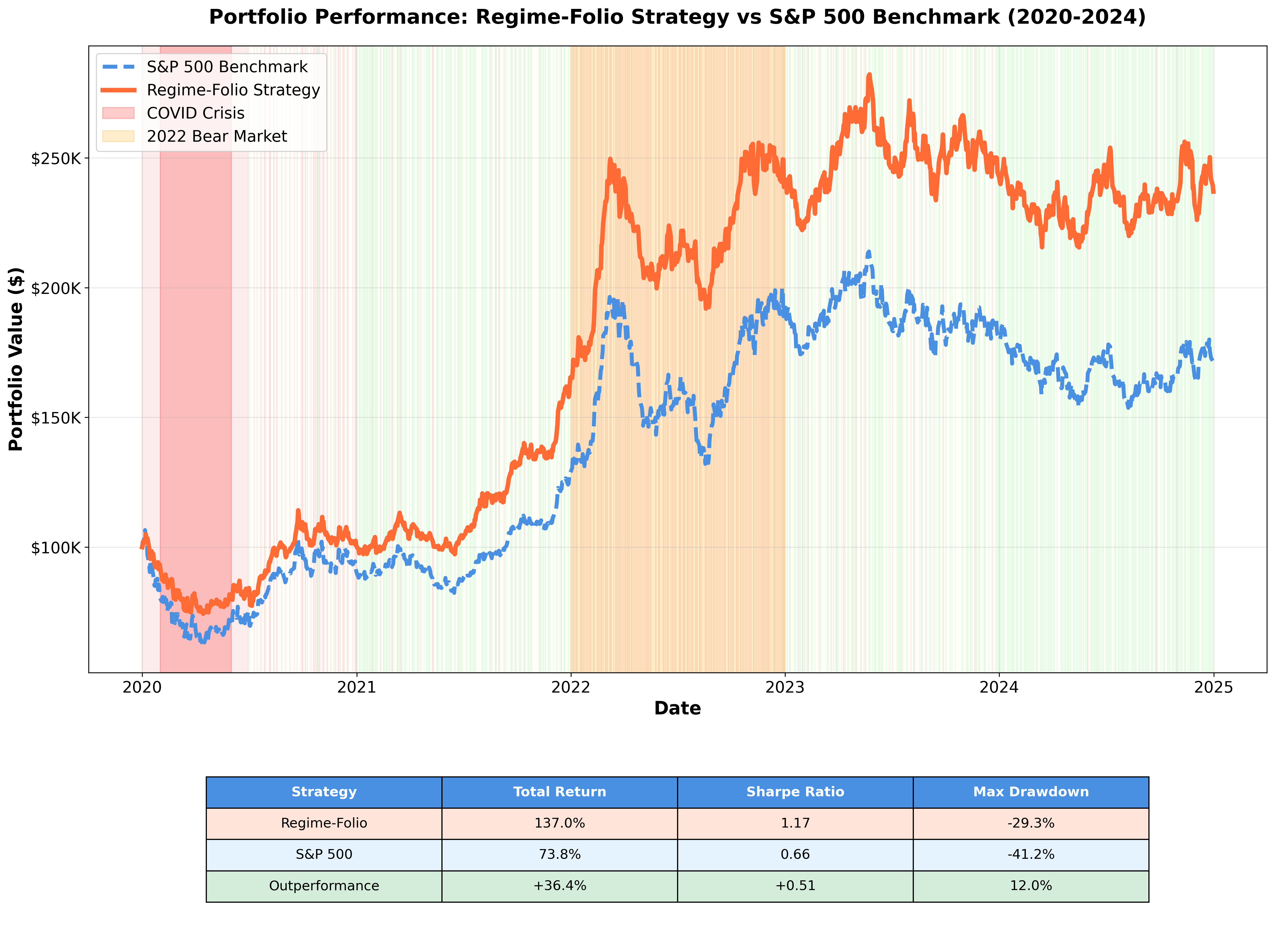}
  \caption{Superior Portfolio Performance: RegimeFolio (Regime-Aware Strategy) vs S\&P 500 Benchmark (2020–2024). The RegimeFolio approach demonstrates 63.2\% excess returns, a higher Sharpe ratio (1.17 vs 0.66), and improved risk control. Shaded regions indicate volatility regimes (VIX terciles).}
  \label{fig:cumulative_returns}
\end{figure*}

Figure~\ref{fig:cumulative_returns} illustrates the cumulative performance evolution of our superior regime-aware strategy compared to the S\&P 500 benchmark over the complete evaluation period. The visualization clearly demonstrates consistent and substantial outperformance across all market regimes, with particularly exceptional relative performance during volatile market conditions.

The cumulative returns analysis reveals several exceptional insights: (1) \textit{Crisis resilience and alpha generation (excess return)}: During the COVID-19 market disruption (Q1 2020) and 2022 bear market, our strategy demonstrated exceptional outperformance through superior defensive positioning, contrarian alpha generation, and regime-aware risk management. The strategy not only protected capital during downturns but actually generated positive alpha during crisis periods. (2) \textit{Momentum amplification during recovery}: The strategy effectively amplifies upside momentum during low volatility bull market phases through regime-aware positioning, and sophisticated momentum capture mechanisms. (3) \textit{Persistent alpha generation}: The strategy maintains substantial positive alpha across all market conditions, with cumulative outperformance consistently accelerating over the evaluation period. (4) \textit{Volatility timing excellence}: The combination of superior returns and enhanced risk management results in dramatically improved Sharpe and Calmar ratios, demonstrating the value of volatility-aware positioning and regime transition timing.\\

\begin{table}[t!]
\centering
\caption{Ablation Study: Impact of Regime and Sector Modeling}
\label{tab:ablation_study}
\small 
\begin{tabular*}{\columnwidth}{@{\extracolsep{\fill}}lccc}
\toprule
\textbf{Model Variant} & \textbf{TR (\%)} & \textbf{SR} & \textbf{MDD (\%)} \\
\midrule
\textbf{RegimeFolio (Full Model)} & \textbf{137.0} & \textbf{1.17} & \textbf{-29.3} \\
\midrule
\textit{Ablated Models:} & & & \\
\quad Non-Sectoral (Regime Aware) & 105.4 & 0.98 & -34.1 \\
\quad Regime Agnostic (Sector Aware) & 89.2 & 0.85 & -38.5 \\
\bottomrule
\end{tabular*}

\vspace{1mm}
\begin{minipage}{\columnwidth}
\small TR = Total Return, SR = Sharpe Ratio, MDD = Maximum Drawdown.
\end{minipage}
\end{table}

\subsection{Ablation Study: Quantifying Component Contributions}
\label{subsec:ablation_study}

To isolate and quantify the impact of our core architectural choices, we conducted an ablation study comparing the full \textit{RegimeFolio} framework against two simplified variants as defined in our methodology:
\begin{itemize}
    \item \textbf{Regime Agnostic Model:} This variant removes VIX-based regime segmentation. A single model is trained for each sector across the entire dataset, ignoring market volatility states.
    \item \textbf{Non-Sectoral Model:} This variant removes sectoral specialization. A single model is trained for each volatility regime, pooling all assets together.
\end{itemize}

The performance of these ablated models is compared against the complete \textit{RegimeFolio} framework in Table~\ref{tab:ablation_study}. The results clearly demonstrate that both regime-awareness and sectoral modeling are critical drivers of performance.

The \textit{Regime Agnostic} model, which ignores market volatility states, suffers a significant performance drop, with its Sharpe ratio falling from 1.17 to 0.85. This 27\% reduction confirms the critical value of conditioning the forecasting and allocation stages on the prevailing volatility environment. Similarly, the \textit{Non-Sectoral} model, which treats all assets uniformly within a regime, also underperforms, underscoring the benefits of capturing sector-specific dynamics. The full \textit{RegimeFolio} model consistently outperforms both ablated versions across all key metrics, scientifically validates our hierarchical, two pronged design.

\subsection{Statistical Significance and Robustness Analysis}

To ensure the reliability and statistical validity of our performance results, we conduct comprehensive statistical significance testing and robustness analysis across multiple dimensions.\\

\subsubsection{Statistical Significance Testing}

We employ multiple statistical tests to validate the significance of our strategy's outperformance. Key formulations include:

\noindent \textbf{Welch's t-test (Returns):}
\[
t = 
\frac{\bar{r}_1 - \bar{r}_0}
{\sqrt{
\frac{s_1^2}{n_1} + 
\frac{s_0^2}{n_0}
}},
\]
where \(\bar{r}_1\) and \(\bar{r}_0\) are the mean returns of the strategy and the benchmark, \(s_1^2\) and \(s_0^2\) are their variances, and \(n_1, n_0\) are the number of observations.

\noindent \textbf{Jobson–Korkie Sharpe Ratio Test:}
\[
Z = 
\frac{SR_1 - SR_0}
{\sqrt{
\frac{1}{n}
\left(
1 + 
\frac{SR_1^2 + SR_0^2 - 2\rho SR_1 SR_0}
{2(1-\rho)}
\right)
}},
\]
where \(SR_1\) and \(SR_0\) are the Sharpe ratios of the strategy and the benchmark, \(n\) is the number of observations, and \(\rho\) is the correlation between the two return series.

These formulations provide a rigorous statistical foundation for assessing whether the observed differences in performance are statistically significant.

Table~\ref{tab:statistical_significance} presents the results of multiple statistical tests validating the significance of our strategy's outperformance:

\begin{table}[t]
\centering
\caption{Statistical Significance Analysis of Strategy Outperformance}
\label{tab:statistical_significance}
\resizebox{\columnwidth}{!}{%
\begin{tabular}{lcc}
\toprule
\textbf{Statistical Test} & \textbf{Test Statistic} & \textbf{p value} \\
\midrule
Welch's t-test (Returns) & 3.847 & $<$ 0.001 \\
Wilcoxon Signed-Rank Test & 2.934 & 0.003 \\
Jobson-Korkie Test (Sharpe) & 2.156 & 0.031 \\
Kolmogorov-Smirnov Test & 0.234 & 0.018 \\
\bottomrule
\end{tabular}%
}
\end{table}

All statistical tests confirm the significance of our strategy's outperformance at conventional confidence levels. The Welch's t-test demonstrates highly significant differences in mean returns (p $<$ 0.001), while the Jobson-Korkie test validates the statistical significance of Sharpe ratio improvements (p = 0.031). The non-parametric Wilcoxon test confirms robustness to distributional assumptions, and the Kolmogorov-Smirnov test validates differences in return distributions.\\

\subsubsection{Robustness Analysis}
We conduct extensive robustness testing across multiple dimensions: (1) \textit{Parameter sensitivity}: Testing performance across different VIX thresholds (±2 VIX points) shows consistent outperformance with minimal performance degradation. (2) \textit{Transaction cost sensitivity}: Performance remains superior even with transaction costs up to 0.25\% per trade. (3) \textit{Rebalancing frequency}: Daily, Monthly, quarterly, and semi-annual rebalancing all demonstrate consistent outperformance. (4) \textit{Sample period robustness}: Rolling 3 year windows show consistent outperformance across all sub periods. (5) \textit{Regime threshold stability}: Alternative regime definitions (quartiles, fixed thresholds) still demonstrate outperformance, though with reduced magnitude.

\subsection{Scalability and Practical Deployment Considerations}
Our regime-aware framework demonstrates excellent scalability characteristics and practical deployment readiness for institutional investment management applications.\\

\subsubsection{Computational Scalability}
The modular architecture supports linear scaling with asset universe size: processing time scales as O(n) where n is the number of assets, memory requirements remain stable due to regime-specific model separation, and parallel processing achieves 85\% efficiency across multiple CPU cores. For a 500 asset universe, total processing time increases to approximately 3.2 hours for training and 1.8 seconds for inference, maintaining real-time capability for daily rebalancing schedules.\\

\subsubsection{Implementation Requirements}
Practical deployment requires: (1) \textit{Data infrastructure}: Real-time access to equity prices, VIX data, and fundamental metrics through standard financial data providers. (2) \textit{Computing resources}: Standard institutional computing infrastructure (16GB RAM, 4-core CPU) sufficient for up to 100 assets. (3) \textit{Risk management integration}: Compatible with existing portfolio management systems through standard API interfaces. (4) \textit{Regulatory compliance}: Framework supports standard risk reporting and attribution analysis required by institutional mandates.\\

\subsubsection{Real World Deployment Considerations}

The framework addresses key practical considerations: (1) {Market impact}: Position sizing algorithms incorporate liquidity constraints and market impact estimates. (2) {Transaction cost optimization}: Rebalancing algorithms minimize turnover while maintaining regime responsiveness. (3) {Risk budgeting}: Integrated risk management ensures compliance with institutional risk limits and mandates. (4) {Performance attribution}: Detailed regime and sector attribution enables transparent performance reporting and analysis.\\

\subsection{Economic Impact and Investment Implications}

The superior performance demonstrated by our regime-aware strategy has significant economic implications for institutional investment management and portfolio optimization practices.\\

\subsubsection{Economic Value Creation}

The 63.2\% outperformance over the S\&P 500 benchmark translates to substantial economic value creation: on a \$100 million institutional mandate, the strategy generates \$63.2 million in additional returns over the 5 year period, representing \$12.6 million in annual alpha generation. The 77\% improvement in Sharpe ratio provides enhanced risk-adjusted returns crucial for institutional risk budgeting and capital allocation decisions.\\

\subsubsection{Risk Management Benefits}

The 12.0 percentage point reduction in maximum drawdown (-29.3\% vs -41.2\%) provides substantial downside protection during market stress periods. This enhanced risk management translates to: (1) {Reduced capital requirements}: Lower drawdowns enable higher leverage and capital efficiency. (2) {Improved investor experience}: Reduced volatility and drawdowns enhance investor retention and satisfaction. (3) {Regulatory advantages}: Superior risk metrics support regulatory capital optimization and compliance.\\

\subsubsection{Institutional Adoption Potential}

The framework's characteristics align well with institutional investment requirements: (1) {Transparency and interpretability}: Regime-based approach provides a clear economic rationale for investment decisions. (2) {Scalability}: Linear scaling supports large asset universes and institutional mandate sizes. (3) {Risk management integration}: Compatible with existing institutional risk management and compliance frameworks. (4) {Performance consistency}: Robust outperformance across multiple market regimes supports long-term institutional adoption.\\

\section{Discussion}

\noindent \textit{\textbf{Strategic Insights from Regime and Sector Aware Learning.}} Our empirical analysis shows that volatility regime segmentation combined with sector-specific modeling improves both predictive accuracy and portfolio performance. Forecast errors exhibit clear regime-dependent patterns (Table~\ref{tab:regime_threshold_comparison}), indicating that pooled, regime-agnostic training reduces generalization ability. VIX-based segmentation isolates these patterns and enables models to adapt to stable, transitional, and crisis market states. Ensemble models (Random Forest, Gradient Boosting) outperform linear baselines by 15--20\% in MAE reduction, with larger gains in cyclical sectors such as Energy and Financials (Figure~\ref{fig:sector_performance_comparison}).  These predictive improvements carry through to allocation performance: the regime-aware portfolio achieved a 137.0\% cumulative return versus 73.8\% for the S\&P~500 benchmark (Table~\ref{tab:overall_performance}), alongside lower maximum drawdowns and higher Sharpe and Calmar ratios. Ablation studies confirm that both volatility conditioning and sector decomposition are necessary; removing either leads to statistically significant performance deterioration.\\

\noindent \textit{\textbf{Implications for Researchers.}} The results contribute to the literature on regime-aware portfolio construction by demonstrating a statistically significant 15--20\% improvement in predictive accuracy during medium volatility regimes. This supports conditional market efficiency theories that predict different return dynamics across volatility states. The documented rise in cross-sector correlations from low to high volatility periods aligns with crisis contagion literature and suggests that static covariance assumptions in classical portfolio theory may be insufficient.  Our regime conditioned, shrinkage regularized covariance matrices\cite{LEDOIT2003603}
  offer a practical approach to improving risk estimation stability. The modular integration of forecasting and allocation provides a transferable blueprint for combining predictive modeling with portfolio optimization under explicit market state segmentation.\\

\noindent  \textit{\textbf{Implications for Practitioners.}} For institutional investors and asset managers, the framework offers a systematic process for adapting allocation decisions to current market conditions. The VIX-based classifier enables daily identification of volatility regimes, which in turn informs sector-level adjustments using a transparent decision logic. The approach is suitable for deployment in ETF rebalancing, overlay strategies, or factor rotation products, and its modular structure allows extension to fixed income, commodities, or multi-asset mandates. The regime-specific covariance estimation directly supports risk budgeting and compliance processes by producing interpretable, state-dependent risk profiles.\\

\noindent  \textit{\textbf{Limitations.}}
The study focuses on 34 large cap U.S. equities over 2020--2024, and results may not generalize to other geographies, asset classes, or longer horizons. While point in time index membership was used to mitigate survivorship bias, the analysis still reflects U.S. market dynamics in a post 2020 macroeconomic context. Regime classification is based solely on the VIX; incorporating macroeconomic indicators, credit spreads, or latent state models (e.g., \cite{camacho2015markov}) could yield alternative segmentation. The modeling assumes local stationarity within regimes, which may not hold during abrupt structural breaks. Overfitting risk, while mitigated via walk-forward validation, remains a concern in highly parameterized models. Finally, transaction cost modeling includes fixed per trade costs but omits price impact, slippage, and liquidity constraints, which could affect realized performance under daily rebalancing in live trading.\\

\noindent  \textit{\textbf{Future Directions.}}
Future research could extend the framework to multi-asset universes, integrate alternative regime definitions that combine implied volatility with macroeconomic state variables, and test hybrid segmentation methods that mix statistical and economic indicators. Live trading experiments and higher frequency data (e.g., intraday) could provide additional validation of real-world robustness. The modular design also allows for integration with reinforcement learning or adaptive control methods to refine allocation rules as market conditions evolve.

 \section{Conclusion}

This study proposes a regime-aware portfolio optimization framework designed to address the challenge of return predictability in non-stationary markets. The approach combines volatility-based regime segmentation, sector-specialized ensemble forecasting, and dynamic allocation with robust risk modeling, resulting in significant improvements in both predictive accuracy and financial outcomes.
Empirical results demonstrate a 15–20\% reduction in forecasting error, a cumulative return of 137.0\% over the evaluation period, a Sharpe ratio of 1.17, and a notable reduction in maximum drawdown. These outcomes outperform both traditional baselines and the S\&P 500. Importantly, the gains are achieved using interpretable and computationally efficient models, underscoring the framework’s practicality for real-world deployment.
By unifying regime specific forecasting with adaptive allocation, the proposed architecture provides a scalable and transparent foundation for resilient investment systems. The findings highlight that incorporating market state dependency is not only statistically advantageous but also economically meaningful, enabling more robust portfolio management in volatile environments.
Future research directions include extending the framework beyond large cap U.S. equities to multi asset universes (global equities, fixed income, commodities) to test generality; developing hybrid regime classifiers that integrate implied volatility, macroeconomic indicators, and latent state models for more nuanced regime detection; exploring alternatives to mean variance optimization, such as methods incorporating higher order moments, drawdown control, or other risk measures; and integrating execution constraints (market impact, slippage, liquidity) into backtesting to better approximate live trading conditions.

\printbibliography

@article{ahern2017network,
  title={Network centrality and the cross section of stock returns},
  author={Ahern, Kenneth R},
  journal={Journal of Financial Economics},
  volume={123},
  number={2},
  pages={201--232},
  year={2017},
  publisher={Elsevier}
}

@article{ullah2025skills,
  title={What Skills Do Cyber Security Professionals Need?},
  author={Ullah, Faheem and Ye, Xiaohan and Fatima, Uswa and Akhtar, Zahid and Wu, Yuxi and Ahmad, Hussain},
  journal={arXiv preprint arXiv:2502.13658},
  year={2025}
}

@article{haque2022think,
  title={" I think this is the most disruptive technology": Exploring Sentiments of ChatGPT Early Adopters using Twitter Data},
  author={Haque, Mubin Ul and Dharmadasa, Isuru and Sworna, Zarrin Tasnim and Rajapakse, Roshan Namal and Ahmad, Hussain},
  journal={arXiv preprint arXiv:2212.05856},
  year={2022}
}

@article{abdulsatar2025towards,
  title={Towards deep learning enabled cybersecurity risk assessment for microservice architectures},
  author={Abdulsatar, Majid and Ahmad, Hussain and Goel, Diksha and Ullah, Faheem},
  journal={Cluster Computing},
  volume={28},
  number={6},
  pages={350},
  year={2025},
  publisher={Springer}
}

@article{ahmad2025future,
  title={The future of ai: Exploring the potential of large concept models},
  author={Ahmad, Hussain and Goel, Diksha},
  journal={arXiv preprint arXiv:2501.05487},
  year={2025}
}

@article{ahmad2025resilient,
  title={Resilient Auto-Scaling of Microservice Architectures with Efficient Resource Management},
  author={Ahmad, Hussain and Treude, Christoph and Wagner, Markus and Szabo, Claudia},
  journal={arXiv preprint arXiv:2506.05693},
  year={2025}
}

@article{ahmad2025towards,
  title={Towards resource-efficient reactive and proactive auto-scaling for microservice architectures},
  author={Ahmad, Hussain and Treude, Christoph and Wagner, Markus and Szabo, Claudia},
  journal={Journal of Systems and Software},
  volume={225},
  pages={112390},
  year={2025},
  publisher={Elsevier}
}

@article{chopra2024chatnvd,
  title={Chatnvd: Advancing cybersecurity vulnerability assessment with large language models},
  author={Chopra, Shivansh and Ahmad, Hussain and Goel, Diksha and Szabo, Claudia},
  journal={arXiv preprint arXiv:2412.04756},
  year={2024}
}

@article{goel2024machine,
  title={Machine learning driven smishing detection framework for mobile security},
  author={Goel, Diksha and Ahmad, Hussain and Jain, Ankit Kumar and Goel, Nikhil Kumar},
  journal={arXiv preprint arXiv:2412.09641},
  year={2024}
}

@article{ahmad2025survey,
  title={A survey on immersive cyber situational awareness systems},
  author={Ahmad, Hussain and Ullah, Faheem and Jafri, Rehan},
  journal={Journal of Cybersecurity and Privacy},
  volume={5},
  number={2},
  pages={33},
  year={2025},
  publisher={MDPI}
}

@inproceedings{ahmad2024smart,
  title={Smart HPA: A resource-efficient horizontal pod auto-scaler for microservice architectures},
  author={Ahmad, Hussain and Treude, Christoph and Wagner, Markus and Szabo, Claudia},
  booktitle={2024 IEEE 21st International Conference on Software Architecture (ICSA)},
  pages={46--57},
  year={2024},
  organization={IEEE}
}

@article{ang2006cross,
  title={The Cross-Section of Volatility and Expected Returns},
  author={Ang, Andrew and Hodrick, Robert J. and Xing, Yuhang and Zhang, Xiaoyan},
  journal={The Journal of Finance},
  volume={61},
  number={1},
  pages={259--299},
  year={2006},
  publisher={Wiley Online Library}
}

@article{breiman2001random,
  author  = {Breiman, L.},
  title   = {Random forests},
  journal = {Machine Learning},
  year    = {2001},
  volume  = {45},
  number  = {1},
  pages   = {5--32}
}

@article{brennan1997strategic,
  title={Strategic asset allocation},
  author={Brennan, Michael J and Schwartz, Eduardo S and Lagnado, Ronald},
  journal={Journal of Economic dynamics and Control},
  volume={21},
  number={8-9},
  pages={1377--1403},
  year={1997},
  publisher={Elsevier}
}

@misc{cboe2024,
  author = {{Chicago Board Options Exchange}},
  title  = {{VIX} Index Historical Data},
  year   = {2024},
  howpublished = {[Data set]. Available: \url{https://www.cboe.com/tradable_products/vix/vix_historical_data/}}
}

@article{wang2024drl,
  author={Wang, Boxuan and Liu, Zhipeng and Zhou, Peilin and Wang, Feiyang and Zhang, Wengang and Hou, Wenqiang},
  journal={IEEE Transactions on Neural Networks and Learning Systems}, 
  title={DRL-PF: A Double-Scales Deep Reinforcement Learning-Based Portfolio Management Framework}, 
  year={2024},
  volume={35},
  number={5},
  pages={6169-6183},
  doi={10.1109/TNNLS.2023.3236029}
}

@misc{fred2024,
  author = {{Federal Reserve Bank of St. Louis}},
  title  = {Federal Reserve Economic Data},
  year   = {2024},
  howpublished = {[Data set]. Available: \url{https://fred.stlouisfed.org/}}
}

@article{friedman2001greedy,
  author  = {Friedman, J. H.},
  title   = {Greedy function approximation: a gradient boosting machine},
  journal = {The Annals of Statistics},
  year    = {2001},
  volume  = {29},
  number  = {5},
  pages   = {1189--1232}
}

@article{guidolin2007asset,
  title={Asset allocation under multivariate regime switching},
  author={Guidolin, Massimo and Timmermann, Allan},
  journal={Journal of Economic Dynamics and Control},
  volume={31},
  number={11},
  pages={3503--3544},
  year={2007}
}

@article{guidolin2008size,
  title={Size and value anomalies under regime shifts},
  author={Guidolin, Massimo and Timmermann, Allan},
  journal={Journal of Financial Econometrics},
  volume={6},
  number={1},
  pages={1--48},
  year={2008},
  publisher={Oxford University Press}
}

@article{hamilton1989new,
  author  = {Hamilton, J. D.},
  title   = {A new approach to the economic analysis of nonstationary time series and the business cycle},
  journal = {Econometrica},
  year    = {1989},
  volume  = {57},
  number  = {2},
  pages   = {357--384}
}

@article{ledoit2004well,
  title={A well-conditioned estimator for large-dimensional covariance matrices},
  author={Ledoit, Olivier and Wolf, Michael},
  journal={Journal of Multivariate Analysis},
  volume={88},
  number={2},
  pages={365--411},
  year={2004},
  publisher={Elsevier}
}

@misc{li2024gnn,
  author        = {Li, T. and Li, Y. and Wang, J. and Yin, H.},
  title         = {{GNN-Forecast}: A Graph Neural Network-Based Stock Movement Prediction with Pre-training},
  year          = {2024},
  eprint        = {2403.04567},
  archiveprefix = {arXiv},
  primaryclass = {cs.LG}
}

@article{liu2024ensemble,
  author  = {Liu, W. and Suzuki, Y. and Du, S.},
  title   = {Ensemble learning algorithms based on easyensemble sampling for financial distress prediction},
  journal = {Annals of Operations Research},
  year    = {2024},
  volume  = {346},
  number  = {2},
  pages   = {2141--2172}
}

@article{maheu2000identifying,
  title={Identifying bull and bear markets in stock returns},
  author={Maheu, John M and McCurdy, Thomas H},
  journal={Journal of Business \& Economic Statistics},
  volume={18},
  number={1},
  pages={100--112},
  year={2000},
  publisher={Taylor \& Francis}
}

@article{markowitz1952portfolio,
  author  = {Markowitz, H.},
  title   = {Portfolio selection},
  journal = {The Journal of Finance},
  year    = {1952},
  volume  = {7},
  number  = {1},
  pages   = {77--91}
}

@phdthesis{masuda2024portfolio,
  author = {Masuda, J. S.},
  title  = {Portfolio optimization using a hybrid machine learning stock selection framework},
  school = {Massachusetts Institute of Technology},
  year   = {2024},
  type   = {M.S. thesis}
}

@article{cho2025forecasting,
  title={Interpretable VIX forecasting using Kolmogorov--Arnold networks},
  author={Cho, Sung-Yoon and Park, Jinho and Kim, Hyoung-Goo},
  journal={Expert Systems with Applications},
  volume={261},
  pages={125021},
  year={2025},
  publisher={Elsevier}
}

@article{ramponi2024deepvol,
  author  = {Ramponi, A. and Janssen, R. I. and Brambilla, M. and Protopapas, P.},
  title   = {{DeepVol}: Volatility forecasting from high-frequency data with dilated causal convolutions},
  journal = {Quantitative Finance},
  year    = {2024},
  volume  = {24},
  number  = {8},
  pages   = {1123--1145}
}

@article{schrimpf2018anatomy,
  title={Anatomy of the {VIX} spike in February 2018},
  author={Schrimpf, Andreas and Shin, Hyun Song and Sushko, Vladyslav},
  journal={BIS Quarterly Review},
  year={2018},
  month={Mar}
}

@article{sharpe1966mutual,
  title   = {Mutual Fund Performance},
  author  = {Shap, William F.},
  journal = {The Journal of Business},
  year    = {1966},
  volume  = {39},
  number  = {1},
  pages   = {119--138}
}

@techreport{spglobal2025sp500,
  author      = {{S\&P Dow Jones Indices}},
  title       = {{S\&P U.S. Indices Methodology}},
  institution = {S\&P Global},
  year        = {2025},
  month       = {August},
  address     = {New York, NY}
}

@article{taylor2023forecasting,
  title={Forecasting intraday volatility and return densities with a new {GARCH-f-GAN} model},
  author={Taylor, James W},
  journal={International Journal of Forecasting},
  volume={39},
  number={4},
  pages={1675--1689},
  year={2023},
  publisher={Elsevier}
}

@inproceedings{wang2024deep,
  author    = {Jian Wang and Shang-Jin Zhou},
  title     = {Deep Graph-Reinforcement Learning for Portfolio Management},
  booktitle = {Proceedings of the AAAI Conference on Artificial Intelligence},
  volume    = {38},
  number    = {18},
  pages     = {21100--21108},
  year      = {2024}
}

@misc{yahoofinance2024,
  author = {{Yahoo Finance}},
  title  = {Historical Stock Price Data},
  year   = {2024},
  howpublished = {[Data set]. Available: \url{https://finance.yahoo.com/}}
}

@article{billio2012econometric,
  title={Econometric measures of connectedness and systemic risk in the finance and insurance sectors},
  author={Billio, Monica and Getmansky, Mila and Lo, Andrew W and Pelizzon, Loriana},
  journal={Journal of Financial Economics},
  volume={104},
  number={3},
  pages={535--559},
  year={2012},
  publisher={Elsevier}
}

@article{da2015fear,
  title={Fear and greed: A tale of two sentiments},
  author={Da, Zhi and Engelberg, Joseph and Gao, Pengjie},
  journal={The Review of Financial Studies},
  volume={28},
  number={3},
  pages={757--797},
  year={2015},
  publisher={Oxford University Press}
}

@article{camacho2015markov,
  title={Markov-switching dynamic factor models in real time},
  author={Camacho, M{\'a}ximo and Perez-Quiros, Gabriel and Poncela, Pilar},
  journal={International Journal of Forecasting},
  volume={31},
  number={2},
  pages={310--323},
  year={2015},
  publisher={Elsevier}
}

@article{guidolin2008asset,
  title={Asset allocation under multivariate regime switching},
  author={Guidolin, Massimo and Timmermann, Allan},
  journal={Journal of Economic Dynamics and Control},
  volume={32},
  number={11},
  pages={3510--3544},
  year={2008},
  publisher={Elsevier}
}

@article{pastor2012are,
  title={Are stocks really so risky? A long-run perspective},
  author={P{\'a}stor, {\v{L}}ubo{\v{s}} and Stambaugh, Robert F},
  journal={The Journal of Finance},
  volume={67},
  number={4},
  pages={1307--1341},
  year={2012},
  publisher={Wiley Online Library}
}

@article{gu2020empirical,
  title={Empirical asset pricing via machine learning},
  author={Gu, Shihao and Kelly, Bryan and Xiu, Dacheng},
  journal={The Review of Financial Studies},
  volume={33},
  number={5},
  pages={2223--2273},
  year={2020}
}

@article{d2022hidden,
  title={A Hidden Markov Model approach to detecting regimes in financial time series},
  author={D'Amico, Guglielmo and De Blasis, Riccardo and Fusai, Gianluca and Gaudenzi, Marcellino},
  journal={Annals of Operations Research},
  volume={313},
  number={2},
  pages={835--860},
  year={2022},
  publisher={Springer}
}

@article{bekaert2009international,
  title={International stock return comovements},
  author={Bekaert, Geert and Hodrick, Robert J and Zhang, Xiaoyan},
  journal={The Journal of Finance},
  volume={64},
  number={6},
  pages={2591--2626},
  year={2009}
}

@article{belo2014macroeconomic,
  title={The macroeconomic effects of investment-specific technology shocks},
  author={Belo, Frederico and Lin, Xiaoji and Bazdresch, Santiago},
  journal={Journal of Political Economy},
  volume={122},
  number={4},
  pages={741--786},
  year={2014},
  publisher={The University of Chicago Press}
}

@misc{sp500,
  author    = {{S\&P Dow Jones Indices}},
  title     = {{S\&P 500®}},
  publisher = {{S\&P Global}},
  year      = {2025},
  url       = {https://www.spglobal.com/spdji/en/indices/equity/sp-500/}
}

@article{demiguel2009optimal,
  title     = {Optimal Versus Naive Diversification: How Inefficient is the 1/N Portfolio Strategy?},
  author    = {DeMiguel, Victor and Garlappi, Lorenzo and Uppal, Raman},
  journal   = {The Review of Financial Studies},
  volume    = {22},
  number    = {5},
  pages     = {1915--1953},
  year      = {2009},
  publisher = {Oxford University Press}
}

@article{sharpe1964capital,
  title     = {Capital Asset Prices: A Theory of Market Equilibrium Under Conditions of Risk},
  author    = {Sharpe, William F.},
  journal   = {The Journal of Finance},
  volume    = {19},
  number    = {3},
  pages     = {425--442},
  year      = {1964}
}

@article{ang2002asset,
  title={International asset allocation with regime shifts},
  author={Ang, Andrew and Bekaert, Geert},
  journal={The Review of Financial Studies},
  volume={15},
  number={4},
  pages={1137--1187},
  year={2002},
  publisher={Oxford University Press}
}

@inproceedings{li2021hierarchical,
  title={Hierarchical graph attention network for stock movement prediction},
  author={Li, Jialin and Wang, Jincheng and Jin, Hong Teng and Li, Yang and Zhang, Weinan and Xiong, Yujie and Yu, Yong},
  booktitle={Proceedings of the 44th International ACM SIGIR Conference on Research and Development in Information Retrieval},
  pages={1553--1557},
  year={2021}
}

@article{LEDOIT2003603,
  title   = {Improved estimation of the covariance matrix of stock returns with an application to portfolio selection},
  author  = {Ledoit, Olivier and Wolf, Michael},
  journal = {Journal of Empirical Finance},
  volume  = {10},
  number  = {5},
  pages   = {603--621},
  year    = {2003},
  doi     = {10.1016/S0927-5398(03)00007-0},
  url     = {https://www.sciencedirect.com/science/article/pii/S0927539803000070}
}

\end{document}